\documentclass[preprint,12pt]{elsarticle}

\usepackage{amssymb}

\usepackage{amsmath}
\usepackage{graphicx}
\usepackage{dcolumn}
\usepackage{bm}
\usepackage{xcolor}
\usepackage{textgreek}
\usepackage{array} 
\usepackage{threeparttable} 
\usepackage{lscape} 

\journal{Materials Today}

\begin{document}

\begin{frontmatter}



\title{Extraordinary physical properties of superconducting YBa$_{1.4}$Sr$_{0.6}$Cu$_3$O$_6$Se$_{0.51}$ in a multiphase ceramic material}


\author[inst1]{V. Grinenko}

\affiliation[inst1]{organization={Leibniz-Institute IFW-Dresden},
            addressline={PF 270116}, 
            city={Dresden},
            postcode={D-01171}, 
            country={Germany}}

\author[inst2]{A. Dudka}

\affiliation[inst2]{organization={Shubnikov Institute of Crystallography of Federal Scientific Research Centre ``Crystallography and Photonics"},
            addressline={Russian Academy of Sciences}, 
            city={Moscow},
            postcode={119333}, 
            country={Russia}}

\author[inst3]{S. Nozaki}

\affiliation[inst3]{organization={Nanotechnology Systems Division, Hitachi High-Tech America, Inc.},
            addressline={2500 NE Century Boulevard}, 
            city={Hillsboro},
            postcode={97124}, 
            state={OR},
            country={USA}}

\author[inst4]{J. Kilcrease}

\affiliation[inst4]{organization={Hitachi High-Tech America, Inc.},
            addressline={}, 
            city={Clarksburg},
            postcode={20871}, 
            state={MD},
            country={USA}}

\author[inst4]{A. Muto}

\author[inst4]{J. Clarke}

\author[inst5]{T. Hogan}

\affiliation[inst5]{organization={Quantum Design},
            city={San Diego},
            postcode={92121}, 
            state={CA},
            country={USA}}

\author[inst6,inst7]{V. Nikoghosyan}

\affiliation[inst6]{organization={Physics Research Institute},
            addressline={National Academy of Sciences}, 
            city={Ashtarak},
            postcode={0203}, 
            country={Armenia}}

\author[inst7]{I. de Paiva}

\affiliation[inst7]{organization={Advanced Physics Laboratory, Institute for Quantum Studies},
            addressline={Chapman University}, 
            city={Burtonsville},
            postcode={20866}, 
            state={MD},
            country={USA}}

\author[inst7]{R. Dulal}

\author[inst7]{S. Teknowijoyo}

\author[inst7]{S. Chahid}

\author[inst7]{A. Gulian\corref{cor1}}
\cortext[cor1]{Corresponding author. \ead{gulian@chapman.edu}}

\begin{abstract}
We report on a novel material obtained by modifying pristine YBCO
superconductor in solid phase synthesis via simultaneous partial
substitution of Ba by Sr and O by Se. The intended stoichiometry was
originally YBa$_{2-x}$Sr$_x$Cu$_{3}$O$_{7-x}$Se$_{x}$ with various values of $x$, among
which $x\approx 1$ yielded the best results. Simultaneous application of EDX
and EBSD confirmed that Se atoms indeed enter the crystalline lattice cell.
The detailed XRD analysis further confirmed this conclusion and revealed
that the obtained polycrystalline material contains 5 phases, with the major
phase ($>$30\%) being a cuprate YBa$_{1.4}$Sr$_{0.6}$Cu$_{3}$O$_{6}$Se$%
_{0.51}$. Following XRD analysis, we applied a newly developed approach
which enabled us to determine the ionic positions and occupations in the
cuprate phase. The obtained superconductor demonstrates unique properties,
including i) two superconducting transitions with $T_{c1}\approx$ 35 K
(granular surface phase) and $T_{c2}\approx$ 13 K (bulk granular phase) -
this granular phase arrangement naturally yields the Wohlleben effect; ii)
reentrant diamagnetism and resistive state; iii) strong paramagnetism with Curie-Weiss behavior (%
$\theta_{CW} \approx$ 4 K) and the ferromagnetic phase overruled by
superconductivity; iv) Schottky anomaly visible in the heat capacity data
and most likely delivered by small clusters of magnetic moments. Thorough
analysis of the heat capacity data reveals a strong-coupling $d-$wave
pairing in its bulk phase (with $2\Delta /T_{c}\approx 5$), and, most
importantly, a very unusual anomaly in this cuprate. There are reasons to
associate this anomaly with the quantum criticality observed in traditional
cuprate superconductors at much higher fields (achievable only in certain
laboratories). In our case, the fields leading to quantum criticality are
much weaker ($\sim $7-9 T) thus opening avenues for exploration of the
interplay between superconductivity and pair density waves by the wider
research community.
\end{abstract}




\end{frontmatter}


\section{Introduction}

With the progress in science and technology the discovery of the phenomenon
of superconductivity was inevitable: sooner or later helium would have been
liquefied, and resistivity of metals been tested at liquid helium
temperatures, which actually has been done by Kamerlingh Onnes in 1911 \cite{Onnes1910,matriconBook}.
The discovery of superconductivity in copper oxide materials was much less
probable since the elemental combination Ba$_{x}$La$_{5-x}$Cu$_{5}$O$%
_{5(3-y)}$ with $x =$ 1 and 0.75, $y >$ 0 cannot be delivered by
technical progress, and is, first of all, the product of extraordinary
brainwork and stubbornness of its creators \cite{Bednorz86}. This work generated more
than 20,000 research articles, the most famous of them related to the
Y--Ba--Cu--O superconductor with transition temperature $T_{c}\approx$ 93 K \cite{Wu87}.
Qualitatively, it is a result of compositional substitute of La by Y,
another representative of the same rare earth metals group in Periodic table
of elements, combined with the certain variation of resultant stoichiometry.
The YBCO was a phenomenal success and it became the most famous
superconductor, since it swiftly crossed the liquid nitrogen barrier at
ambient pressure, simplifying greatly the research in novel superconducting
materials and delivering a great puzzle for theoreticians to decipher its
mechanism. This finding stimulated many other cationic substitutions, which
eventually delivered higher transition temperature superconductivity in
compositions Tl--Ca/Ba--Cu--O \cite{Sheng88}, Hg--Ba--Ca--Cu--O \cite{Schilling93}, 
and Bi--Sr--Ca--Cu--O \cite{Maeda88} while eliminating the rare earth element in them.

Anionic substitutions in copper oxides are much less explored
experimentally, though theoreticians have predicted interesting features
(see, e.g., \cite{Yee2014,Yee2015}). Meanwhile, oxygen plays a crucial role in the copper
oxide superconductivity. Recognizing importance of oxygen, there have been
numerous attempts \cite{Palhan1988,Cloots1991,Cooke1999,Gagnon1989,Kambe1988,Yakinci2013,Slebarski1990, Felner1987,Felner1988} to partially replace oxygen by other
representatives (namely, S and Se) of the chalcogen group to which O belongs
with contradicting statements. While successful chalcogen substitution is
claimed in Refs. \cite{Palhan1988,Cloots1991,Cooke1999,Felner1987,Felner1988}, the opposite (no entrance of S into
the lattice) is reported in \cite{Gagnon1989}. Moreover, in another work \cite{Yakinci2013} it is
claimed that Se goes into Y's position. Thus, in view of this ambiguity, the
first task pursued in our research is to prove that Se indeed
enters the crystalline structure and occupies certain O positions in the
lattice. This task is complicated by the fact that the polycrystalline
material under investigation is heterophase: as the analysis revealed, 5
phases are present. Nevertheless, by combining simultaneous EDX-EBSD
analysis with newly suggested method of XRD analysis we were lucky to resolve
this task with accuracy leaving no doubts on the validity the structural model. 
The crystalline lattice parameters, as well as the positions of
Se in the crystalline lattice of superconducting phase, are identified.
Details of our approach are given in Methods. 
Comparing our results with the previous reports, one can conclude that the main
reason for success with chalcogen substitution is in simultaneous partial substitution
of Ba by Sr - this kind of double substitution was not exercised in the
past. The second part of our report addresses the unique physical properties of 
superconducting phase of YBa$_{1.4}$Sr$_{0.6}$Cu$_3$O$_6$Se$_{0.51}$ based on standard set of measurements in 
these multiphase samples. Temperature dependence of resistance, magnetism and heat
capacity reveal two superconducting transitions with $T_{c1} \approx$ 35 K
and $T_{c2} \approx$ 13 K. Above $T_{c1}$, anomalous paramagnetism is
detected with the Curie-Weiss behavior and positive Curie temperature $\theta%
_{Curie} \approx$ 4 K. While no ferromagnetism is visible in DC and AC
magnetization, presence of magnetic moments (caused by changing the Cu-ion
ligands from O to Se) can be deduced from these Curie-Weiss data. The
samples demonstrate reentrant diamagnetism (the Wohlleben effect) whose
mechanism can be explained in combination with other measured quantities.
The heat capacity studies allowed us to conclude that the lower $T_{c}$%
-phase, which constitutes the majority (volume phase), is of the d-wave
nature. Very interestingly, the electronic heat capacity demonstrates
features which may be associated with the quantum criticality border
crossing induced by magnetic field. Observed fluctuations in the AC magnetic
susceptibility further confirm this conjecture. Magnetic field-induced
quantum phase transitions, as well as quantum criticality in general [18],
are a subject of intense research since they are related with the mechanism
of high-temperature superconductivity. Thermodynamic evidence of possible
field-induced border-crossing of quantum criticality in oxy-chalcogen YBa$_{2-x}$Sr$_x$Cu%
$_{3}$O$_{7-x}$Se$_{x}$ superconductor constitutes the third main part of
our research. Importantly, it occurs at much lower than usual magnetic
fields \cite{Shibauchi2008,Michon2019}, and thus opens feasibility avenues for the research community.

\section{Results}

\subsection{Composition of samples}
Our results are obtained on polycrystalline ceramic samples. Initial 
stoichiometries with $x=0$, 0.5, 0.75 and 1 were explored. The most drastic 
changes compared to $x=0$ took place in samples with $x=1$, which will 
be solely described in this article. 

About a dozen
pellets were prepared quite reproducibly (the details are presented in
Methods). Their morphology is shown in Fig. 1.
\begin{figure}
    \centering
    \includegraphics[width=\linewidth]{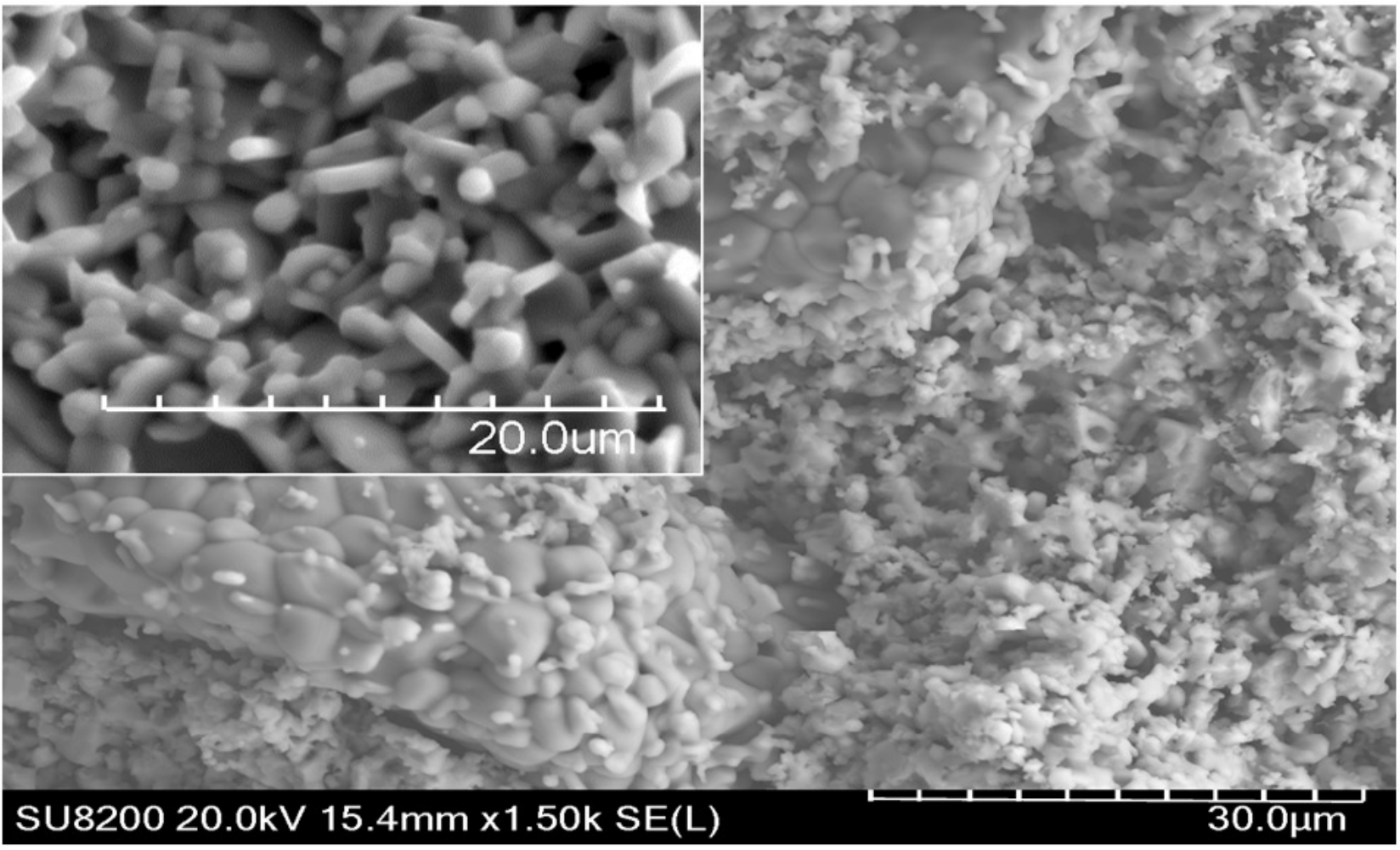}
    \caption{Morphology of freshly
cleaved YBa$_{2-x}$Sr$_x$Cu$_{3}$O$_{7-x}$Se$_{x}$ sample (by Hitachi SU8230 SEM). Inset
- surface pattern (by Hitachi SU3500 SEM) of as-prepared ceramic sample.}
    \label{Fig1}
\end{figure}
The difference in crystallinity of bulk and surface areas is evident from
this figure. Moreover, the heterophase nature is evident in the bulk of the
sample. The phase content would be one of the topics of performed
exploration. However, the first task is to determine the selenium content
in the material. This is a critical topic since the material was synthesized
at high-temperatures (see Methods), and the ionic radius of Se
being higher than O spreads doubts that it may be emanated from the
material. Qualitative confirmation of Se presence was provided by TEM EDX (Fig. 2).

\begin{figure*}
    \centering
    \includegraphics[width=\linewidth]{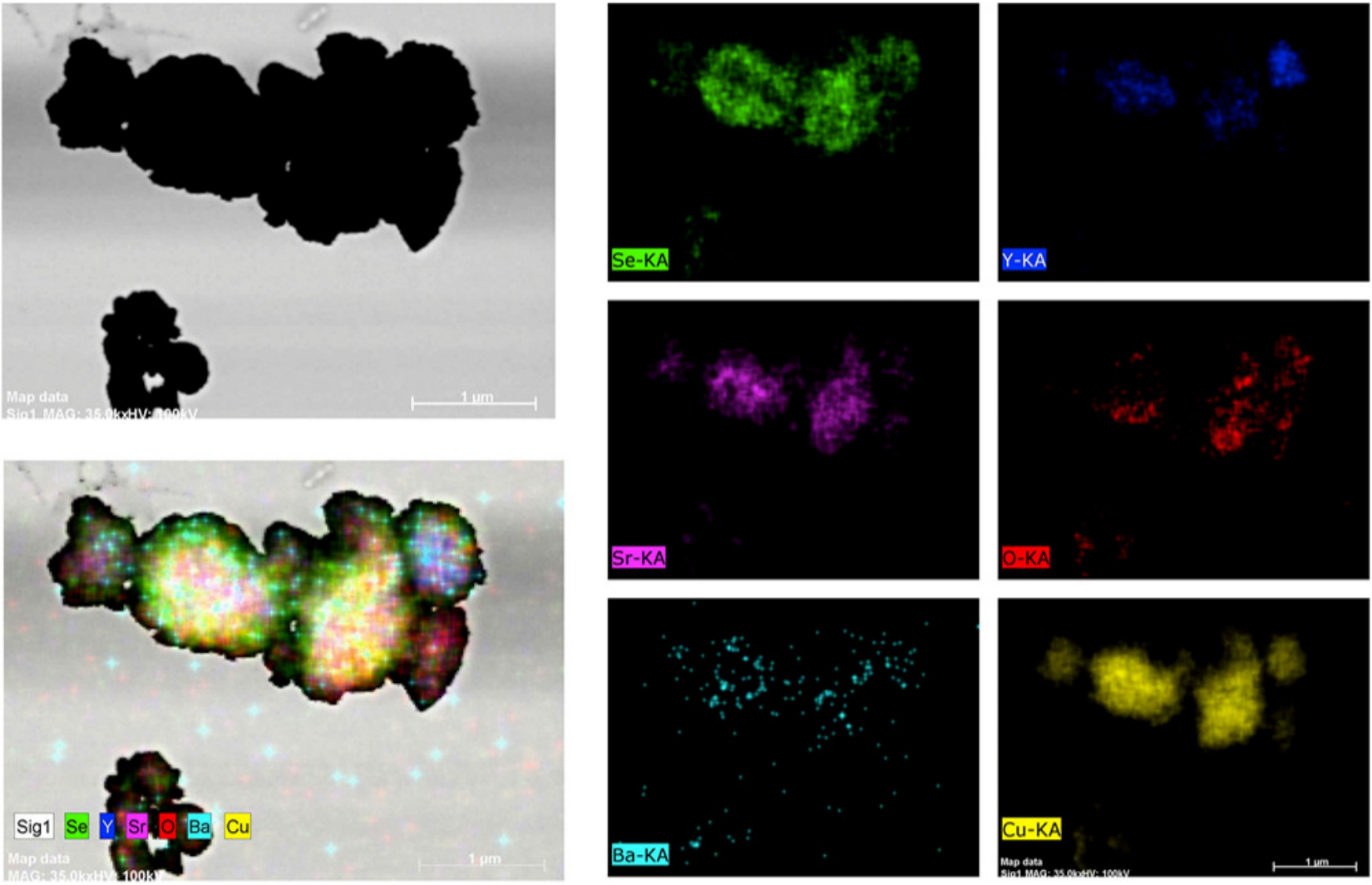}
    \caption{Abundance of Se and other constituent elements (Y, Ba, Sr, O and Cu)
in thin flakes of the powdered polycrystalline YBa$_{2-x}$Sr$_x$Cu$_3$O$_{7-x}$Se$_x$ material with the projected stoichiometry
YBaSrCu$_3$O$_6$Se (transmission EDX by Hitachi HT7700
TEM with Bruker X-flash 6T$|$60 system).}
    \label{Fig2}
\end{figure*}

Quantitative EDX exploration requires flat surfaces of the material, which
is not the case for the flakes in Fig. 1. For that purpose special
techniques were used (see Methods) which yielded the results shown
in Table 1.

\begin{table}
\centering
\caption{EDX data}
\begin{tabular}{|lcccccc|}
\hline
{Elements \%(at)} & {O} & {Cu} & {Se} & {Sr} & 
{Y} & {Ba} \\ 
\hline
{Annealed surface} & 50.2 & 23.1 & 0.3 & 9.0 & 9.0 & 8.4 \\ 
\hline
{Cleaved surface} & 55.5 & 20.6 & 1.0 & 7.0 & 8.6 & 7.3 \\ 
\hline
{Ion polished surface} & 47.0 & 24.7 & 4.9 & 6.9 & 7.7 & 9.0 \\ \hline
\end{tabular}
\end{table}

The most accurate, last row in Table 1, is encouraging; however, it does not
yet guarantee that Se is hosted in any of the unit cells of crystalline
phases. Two more steps were taken for addressing this topic. The first of
them consisted of simultaneous EBSD and EDX analysis, and the second - in
the XRD exploration. 

For the simultaneous EBSD and EDX analysis, the
electron beam was positioned on a crystallite with magnification up to $%
\times $200,000 (Fig. 3\textbf{a}) so as the Kikuchi pattern is revealed
(Fig. 3\textbf{b}) and the corresponding phase is identified (Fig. 3\textbf{c}). 
Without changing the position of the beam the corresponding EDX spectrum
is acquired (Fig. 3\textbf{d}).
\begin{figure*}
    \centering
    \includegraphics[width=\linewidth]{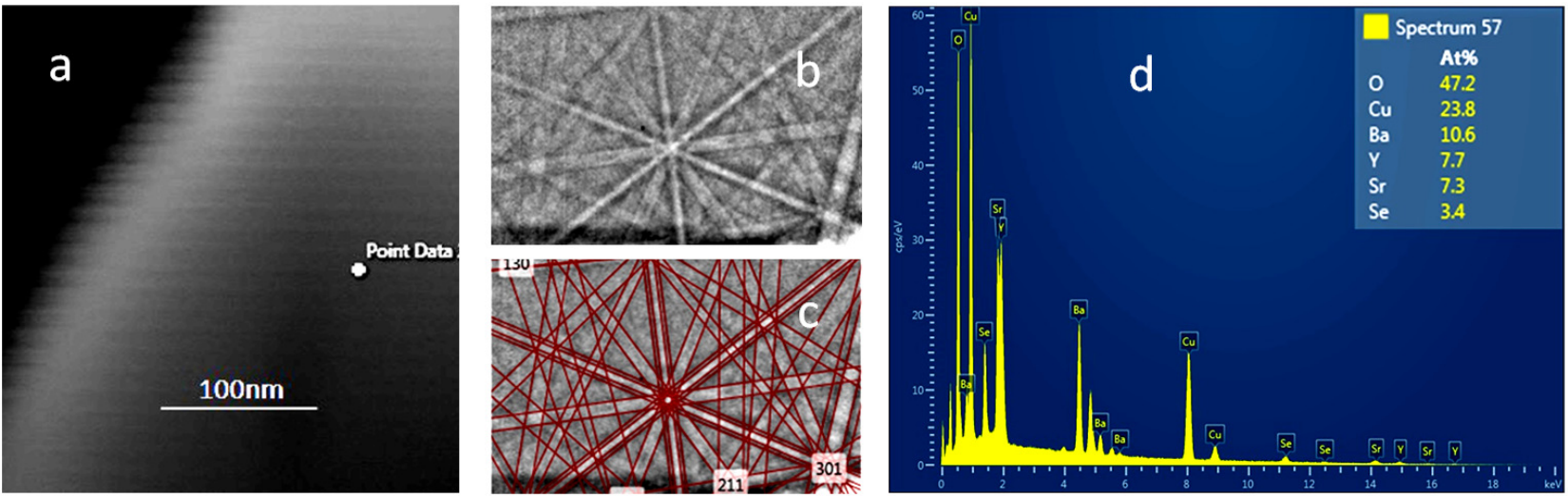}
    \caption{Compositional characterization of polycrystalline samples with the
initially intended stoichiometry YBaSrCu$_{3}$O$_{6}$Se via simultaneous EBSD/EDX
analysis. (\textbf{a}) Material's granule at $\times $200,000 magnification
with indication of spot to which the electron beam was focused. (\textbf{b})
Sample orientation revealed an EBSD Kikuchi patern. (\textbf{c}) EBSD
database recognition of Kikuchi pattern. (\textbf{d}) EDX microanalysis at
the same spot.}
    \label{Fig3}
\end{figure*}
Since the electron beam focusing spot is about 5 nm, and the phase
identified in Fig. 3\textbf{c} corresponds to the YBa$_{2}$Cu$_{3}$O$_{7}$
structure, one can deduce, in view of composition shown in Fig. 3\textbf{d},
that most likely Se enters the crystalline cells of the desired phase YBaSrCu%
$_{3}$O$_{6}$Se. This needs a confirmation.

X-ray diffraction indeed confirms this conclusion and
provides very valuable information about the position of atomic constituents
in the crystalline lattice. The results of XRD analysis are shown in Fig. 4
and Tables 2 and 3 (with more details given in Methods).

\begin{figure}
    \centering
    \includegraphics[width=0.7\linewidth]{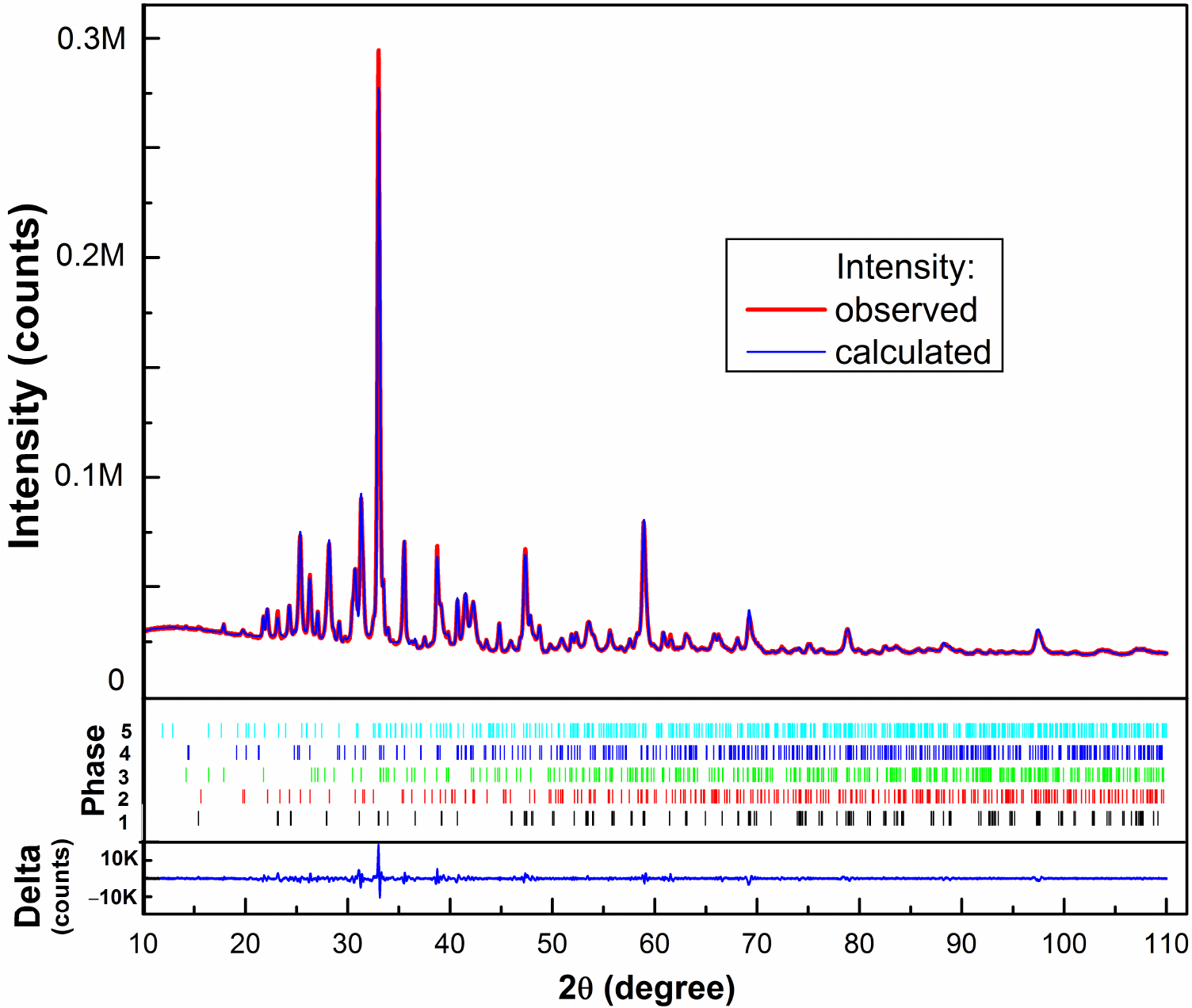}
    \caption{X-ray diffractogram of heterophase ceramic material with the initially
intended stoichiometry YBaSrCu$_{\mathrm{3}}$O$_{\mathrm{6}}$Se. All the peculiarities 
of experimental profile are explained by reflections generated by 5 
phases (see Table 2 and more details in Methods).}
    \label{Fig4}
\end{figure}

After performing Rietveld refinement, the ratio of phase contribution to
X-ray scattering was identified as 30.6 : 26.0 : 24.9 : 13.7 : 4.8 \textit{at.}\% 
for phases 1 through 5 shown in Table 2.

\begin{table*}
\centering
\caption{The comparison between the literature and observed values for the
proposed phases.}
\resizebox{\linewidth}{!}{
\begin{tabular}{|ccllcclll|}
\hline
{Phase} & \%  &{Space group, No} & {Chemical formula} & {Source} & {ICSD} & \textit{a}, \AA  & \textit{b}, 
\AA  & \textit{c}, \AA  \\ \hline
1& 30.6 & $Pmmm$, 47 & YBa$_{1.4}$Sr$_{0.6}$Cu$_{3}$O$_{6}$Se$_{0.51}$ & \cite%
{Licci1998} & 87098 & 3.80174(12) & 3.85009(14) & 11.5739(4) \\ 
& & &  & Our work &  & 3.84359(4) & 3.83295(6) & 11.47711(18) \\ \hline
2& 26.0 & $Pnma$, 62 & BaSeO$_{4}$ & \cite{Pistorius1962} & - & 8.9461(6) & 
5.6911(4) & 7.3313(5) \\ 
& & &  & Our work &  & 8.8970(2) & 5.6891(2) & 7.30815(18) \\ \hline
3& 24.9 & $Pna2_{1}$, 33 & Y$_{2}$Cu$_{2}$O$_{5}$ & \cite{Adams1992,Aride1989} & 
067119,072058 & 10.8038(6) & 3.49600(18) & 12.4665(7) \\ 
& & &  & Our work &  & 10.78668(19) & 3.49389(5) & 12.4404(3) \\ \hline
4& 13.7 & $Pnma$, 62 & Y$_{2}$BaCuO$_{5}$ & \cite{Hsu1996} & 063425 & 12.1793(7) & 
5.6591(5) & 7.1323(4) \\ 
& & &  & Our work &  & 12.30548(14) & 5.67355(14) & 7.03737(13) \\ \hline
5& 4.8 & $Pmn2_{1}$, 31 & BaCu(SeO$_{3}$)$_{2}$ & \cite{Effenberger1987} & 202386
& 13.353(2) & 5.247(1) & 8.981(1) \\ 
& & &  & Our work &  & 13.735(3) & 5.3999(14) & 8.867(2) \\
\hline
\end{tabular}}
\end{table*}

Phase 1, YBa$_{1.4}$Sr$_{0.6}$Cu$_{3}$O$_{6}$Se$_{0.51}$, is basic for this research.
Substantiation of the choice of one of three most probable models and
consecutive analysis of them are described in Methods. The
improvement of structural analysis criteria (such as $R$-factors, 
\textit{R}1/\textit{wR}, the extrema of difference Fourier synthesis,
$\Delta \rho _{\max }/\Delta \rho _{\min }$) takes
place with each next model. Model 1: no Se in any oxygen position \cite%
{Licci1998}. From this, we move to Model 2, where Se partially replaces Y 
\cite{Yakinci2013}. Finally, we deal with Model 3, in which Se partially
replaces O. For these modeling steps, we found consecutively $R1/wR = 0.98/0.90 
\rightarrow  1.09/0.93\rightarrow  0.87/0.73$ and $\Delta \rho _{\max
}/\Delta \rho _{\min }$ = +0.30 / -0.48 $\rightarrow $+0.36 / -0.54 $%
\rightarrow $ +0.17 / -0.30 e/\AA $^{3}$. Final model has common criteria
for accuracy based on 20001-point profile expressed by $R_{p}$ = 1.32\%, $wR%
_{p}$ = 1.92\%, \textit{GOF} = 3.14, which confirms high enough reliability of the
performed structural analysis \cite{Toby2006}. The relevant crystalline
parameters are given in Table 3.

\begin{table*}
\centering
\resizebox{\linewidth}{!}{
\begin{threeparttable}
\caption{{Wyckoff} positions, occupancies (\textit{q}), atomic coordinates (\textit{x/a}, \textit{y/b},
\textit{z/c}), and equivalent isotropic displacement parameters (\textit{U}, \AA $^{2}$) for
the basic atoms in the crystal structure of YBa$_{1.4}$Sr$_{0.6}$Cu$_{3}$O$_{6}$Se$_{0.51}$}
\begin{tabular}{|cclllll|}
\hline
Atom & {Wyckoff} position & \textit{q} & \textit{x/a} & \textit{y/b} & \textit{z/c} & \textit{U} \\ \hline
Y1 & 1\textit{h} & 1 & 0.5 & 0.5 & 0.5 & 0.02303(6) \\ \hline
Ba1\tnote{a} & 2\textit{t} & 0.702(1) & 0.5 & 0.5 & 0.183335(4) & 0.022926(8)
\\ \hline
Sr1 & 2\textit{t} & 0.298(1) & 0.5 & 0.5 & 0.183335(4) & 0.022926(8) \\ \hline
Cu1 & 1\textit{a} & 1 & 0 & 0 & 0 & 0.03563(10) \\ \hline
Cu2 & 2\textit{q} & 1 & 0 & 0 & 0.348091(19) & 0.03616(7) \\ \hline
O1 & 2\textit{q} & 1 & 0 & 0 & 0.11896(5) & 0.0281(4) \\ \hline
O2 & 2\textit{s} & 1 & 0 & 0 & 0.40942(7) & 0.0203(2) \\ \hline
O3 &  2\textit{r} &  1  &  0  &  0.5  &  0.35568(8)   &   0.0346(3) \\ \hline 
Se4 & 2\textit{k} & 0.1628(5) & 0.0391(8) & 0 & 0 & 0.0121(3) \\ \hline
Se5 & 1\textit{b} & 0.1872(8) & 0.5 & 0 & 0 & 0.0049(4) \\
\hline
\end{tabular}
\begin{tablenotes}
     \item[a] {Separate refinement of parameters Ba1 and Sr1 in the full-matrix
mode is possible and stable (the Sr1 atom shifts down along the coordinate \textit{z}
at a certain increase of its parameter \textit{U}). In the current case, we preferred
avoiding insertion of additional refinement parameters in favor of keeping
the larger ratio of the measurements to the number of parameters.}
\end{tablenotes}
\end{threeparttable}}
\end{table*}

Possible contribution of other phases is mentioned in Discussion.

\subsection{Basic superconducting properties.}
Figure 5 reveals magnetic and resistive properties of the samples.

\begin{figure*}
    \centering
    \includegraphics[width=\linewidth]{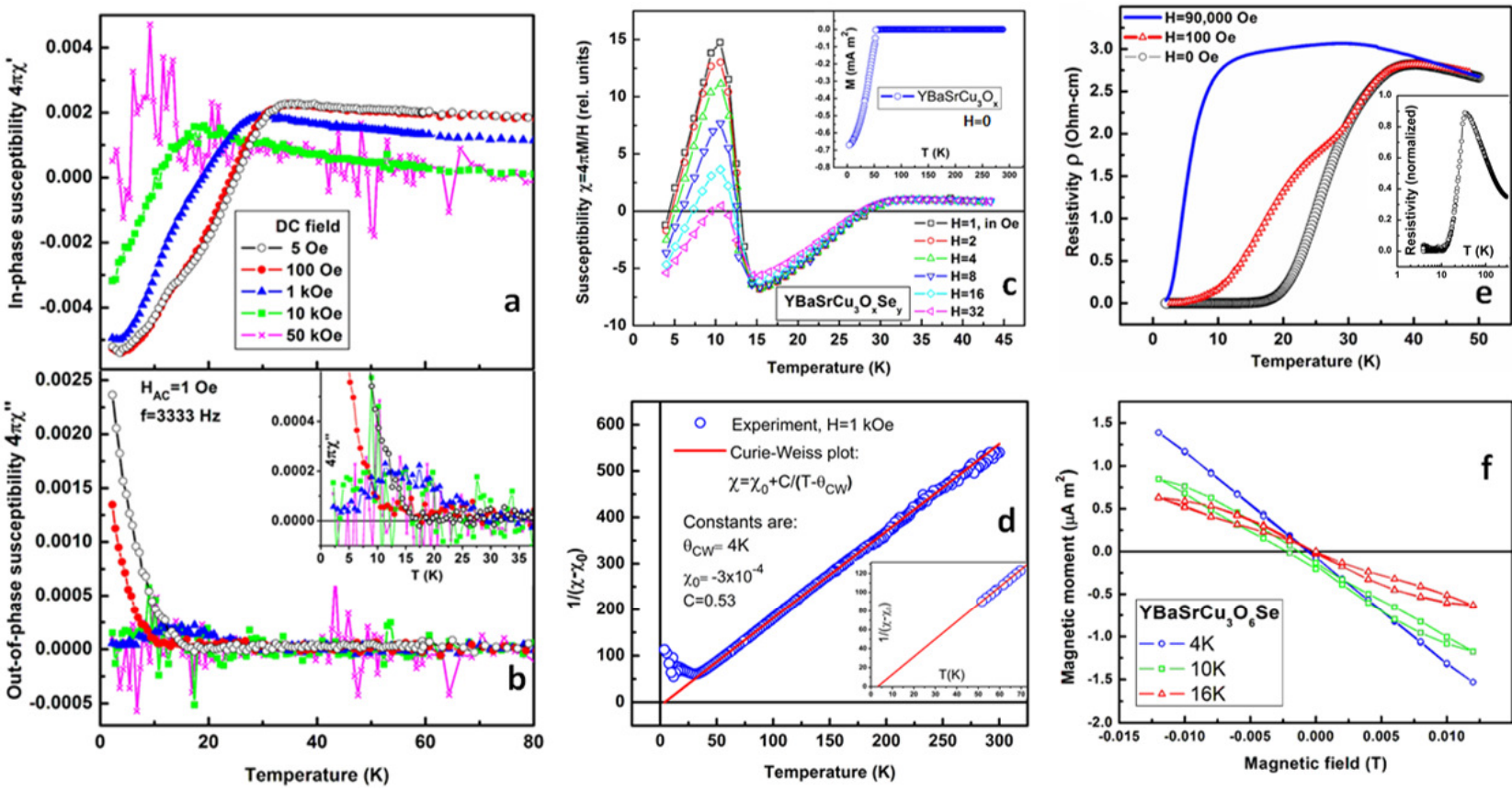}
    \caption{Sample properties in AC and DC magnetic fields. (\textbf{a}) Real
and (\textbf{b}) imaginary parts of AC susceptibility (at $f=$3333 Hz and $%
H_{AC}$=1 Oe) are shown in various DC magnetic fields from 5 Oe to 50 kOe;
inset in (\textbf{b}) illustrates the transition in greater detail. (\textbf{%
c}) Peculiar behavior (reentrant diamagnetism) of DC susceptibility with a
strong paramagnetism above superconducting transition yielding the
Curie-Weiss behavior; inset shows similar transition in a sample without Se.
(\textbf{d}) Curie-Weiss behavior of experimental data on paramagnetism;
inset shows linear extrapolation of experimental data in greater detail
(circles correspond to average of 2 measurements). (\textbf{e})
Magnetoresistance in weak and strong magnetic fields. 
Inset shows noticeable re-entrant resistivity at $T \lesssim4$ K (see also \cite{Gulian2015}).
(\textbf{f}) $\mathbf{M(H)}$ at three characteristic temperatures.}
    \label{Fig5}
\end{figure*}

Observations shown in Fig. 5(\textbf{a}-\textbf{f}) are complemented by the
heat capacity measurements, Fig. 6.

\begin{figure*}
    \centering
    \includegraphics[width=\linewidth]{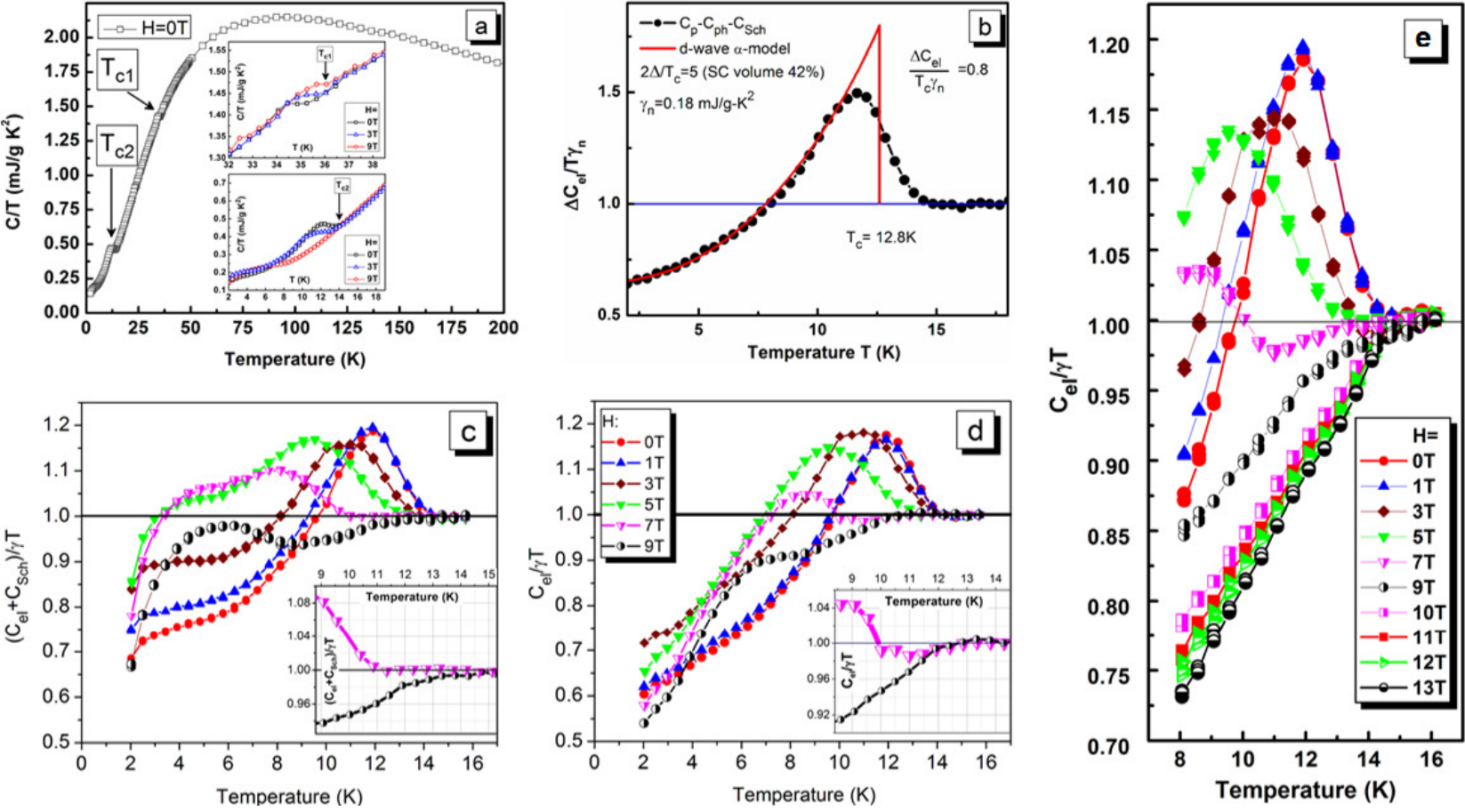}
    \caption{Heat capacity of the sample with the initially intended stoichiometry YBaSrCu$_{3}$O$_{6}$Se. (\textbf{%
a}) Two superconducting transitions (with T$_{c1}$ and T$_{c2}$) at $H=0$;
insets illustrate details of these transitions in $H=0$, 3 and 9 T fields. (%
\textbf{b}) Modeling the transition of the major superconducting phase by
the $d-$wave $\alpha -$model at $H=0$. (\textbf{c}) Behavior of electronic
specific heat with phonon contribution subtracted and Schottky anomaly
present at bulk superconducting transition in different magnetic fields (the
curve coding is similar to panel \textbf{d}). \textbf{(d)} The same behavior with the Schottky anomaly subtracted. Insets in panels \textbf{(c)} and \textbf{(d)} clarify the downturn at $H= 9$ T. \textbf{(e)} Specific heat anomaly with increasing magnetic field (from 0 to 13 T) which reveals quantum criticality (explained in Discussion).}
    \label{Fig6}
\end{figure*}

The data presented in Fig. 5 and Fig. 6 provide very fertile grounds for
discussion.

\section{Discussion}

\subsection{Diamagnetism}
The AC susceptibility $\chi ^{\prime }(T)$ (Fig. 5\textbf{a}) clearly
indicates superconducting transition at about 30 K. More sophisticated is
the behavior of the DC susceptibility (Fig. 5\textbf{c}) which reveals a
peculiarity known as the Wohlleben effect (it will be discussed after
Magnetoresistivity to better understand the material). Despite positiveness
of DC susceptibility at certain temperature range, the dependence of
magnetic moment vs. applied field, \textbf{M(H)} (Fig. 5\textbf{f}), has negative
derivatives at all temperatures below $T_{c}$ $-$ a feature which
corresponds to the superconducting state.

\subsection{Magnetoresistivity}
The second superconducting phase, which is responsible for the reentrant
diamagnetism in Fig. 5\textbf{c}, is also noticeable in magnetoresistivity
(Fig. 5\textbf{e}). This magnetoresistive behavior has a remarkable
low-field feature ($H=0$ Oe and 100 Oe in Fig. 5\textbf{e}): above a certain
temperature $T>T_{0}$, $\rho(T)(H=0)$ and $\rho(T)(H=100)$ match with
one another; however, at $T<T_{0}$, the higher-field curve delays with the
transition to $R=0$ state. This is an unusual feature, caused, most likely,
by the electron hopping. At lower temperatures and higher fields, the
hopping through the grain boundary barrier becomes prohibited (or,
alternatively, charge carrier trapping becomes more effective) yielding this
observation. In stronger fields, the anomaly is gone and the $\rho (T)$
behavior becomes typical, as Fig. 5\textbf{e} demonstrates at $H=90$ kOe.
The divergent behavior of $\chi ^{\prime \prime }$ (Fig. 5\textbf{b}) is
also most likely related to this mechanism: superconductivity in the
intergranular areas disappears in higher fields, leaving only bulk
intragranular effects both in $\chi ^{\prime }$ and $\chi ^{\prime \prime }$. 
Notably, at $H=50$ kOe, the fluctuations in the AC susceptibility become
large and mask off the superconducting transition.

\subsection{Paramagnetism and Curie-Weiss behavior}
The comparison of major curve in Fig. 5\textbf{c} with its inset indicates strong
paramagnetism. Fitting the experimental data above $T_{c1}$ in YBa$_{2-x}$Sr$_x$Cu$_{3}$O%
$_{7-x}$Se$_{x}$ by the Curie-Weiss law $\chi =\chi _{0}+C/(T-\theta _{CW})$%
, as shown in Fig. 5\textbf{d}, one can get the Curie-Weiss temperature $%
\theta _{CW}\approx 4$ K at parameters $\chi _{0}\approx -3\times 10^{-4}$
and $C\approx 0.53$. This implies \cite{Levy1968,Grinenko2011}:%
\begin{equation}
\frac{\mu _{\mathrm{eff}}}{\mu _{B}}\equiv p_{\mathrm{eff}}=\sqrt{C}\frac{%
\sqrt{3k_{B}}}{\mu _{B}\sqrt{yN_{A}}}\approx 2.8\sqrt{\frac{C}{y}}\approx 2.8
\label{1}
\end{equation}%
where $y$\ is the actual Se content in YBa$_{2-x}$Sr$_x$Cu$_{3}$O$_{7-x}$Se$_{x}$ (which,
as follows from Table 1 and the XRD data, is about $0.5$ $-$ the number used
in (\ref{1})). This means that per unit cell the average magnetic moment is $%
\mu _{\mathrm{eff}}\approx 1.5\mu _{B}$ (since only $50\%$ of the
superconducting phase possess Se). Also, one can estimate the total angular
momentum $J$ from the expression \cite{Grinenko2011}:%
\begin{equation}
\sqrt{J(J+1)}=p_{\mathrm{eff}}/g\approx 1.4  \label{2}
\end{equation}%
as $J=1$ (we adopted the electron gyromagnetic ratio $g=2$). Presence of
positive Curie-Weiss temperature typically indicates ferromagnetism at $%
T<\theta _{CW}$. However, as follows from Fig. 5a, there is no ferromagnetic
response visible in $\chi ^{\prime }$ below $T=4$ K which is likely due to
the competing superconducting order\footnote{This fact also indicates that the Curie-Weiss behavior is a property of the
superconducting phase in our heterophase sample.}. However, small
ferromagnetic clusters are most likely existent, which can explain the small
upturn (reentrant resistivity) visible at $T\approx 4$ K 
in Fig. 5\textbf{e} inset. The upturn is also noticeable in $\chi'$ (Fig. 5\textbf{a})
in small magnetic fields, $H=5$ Oe and $H=100$ Oe.
With the field increase, at $H=0.1$ T, the upturn becomes less steep, and
completely disappears in relatively strong fields ($H=1$~T and $H=5$ T).
This can be explained by the orientation of magnetic moments of clusters,
which makes the interaction of these moments with the external field less
effective than for the random orientation. It points towards a preferable
angle of orientation (easy axis) for the AC-field interaction with the
magnetic moments, i.e., towards possible anisotropy. If this conclusion is
correct, then much stronger upturns could appear at certain orientations of
the applied magnetic field relative to the wave vector of the AC-probe
field. Testing of this possible anisotropy could be considered as an
interesting topic, though it is beyond the scope of the present article.

\subsection{Wohlleben effect}
Numerous explanations have been suggested for understanding the mechanism of
curves similar to ones shown Fig. 5\textbf{c} since the first discovery in
granular high-temperature superconductors \cite{Yeshurun1987,Svendlindh1989}%
, and later in many other superconducting objects including Al \cite%
{Geim1998} and Nb \cite{Palacios1998} disks, Josephson junction systems \cite%
{AraujoMoreira1997,Barbara1999}, In-Sn spheres \cite{Chu2006}, etc.

From a theoretical point of view, the magnetic moment (or magnetization 
$\mathbf{M=B-H}$) of a superconductor is determined by a superfluid current $\mathbf{j}_{s}$:
\begin{equation}
\mathbf{M}=\frac{1}{2c}\int \left[ \mathbf{j}_{s}\times \mathbf{r}\right] dV.
\label{A1}
\end{equation}
Thus, theoretical attempts to explain the effect ought to
modify the traditional, Meissner-type behavior of these currents $\mathbf{j}%
_{s}(\mathbf{r},t)$ in a drastic manner to switch from the diamagnetic to
the paramagnetic response. These attempts can principally be classified into
two categories of models, the first of which is based on currents
spontaneously flowing in superconductors in absence of an external field 
$\mathbf{H}_{ext}$. This field orients the randomly distributed local moments $\mathbf{M}_{i}$
so that $\left\langle \mathbf{M}_{i}\right\rangle >0$. The second category
of models depends on more geometrically complex current patterns (in
presence of $\mathbf{H}_{ext}$) than envisioned by the ordinary theoretical approach.
Because of the granular structure (Fig. 1), as well as the presence of
effective magnetic moments in the lattice at low temperatures, the $\pi -$%
junction model \cite{SigristRice1995} was a direct candidate to consider in
our case. The likelihood of applicability of this model was further enhanced
by the notion (discussed later in this article) that the $d-$wave
superconductivity fits best to our heat capacity data (Fig. 6\textbf{b}). To
judge the applicability of the $\pi -$junction model, we demolished one of
our samples and ground it down to submicron sizes by milling it in a
sapphire mortar. The powder was then thoroughly mixed with a glue, so as to
eliminate intergranular couplings. Figure 7 compares the resultant DC
susceptibility with that of the sample while still in a ceramic form.

\begin{figure}
    \centering
    \includegraphics[width=0.6\linewidth]{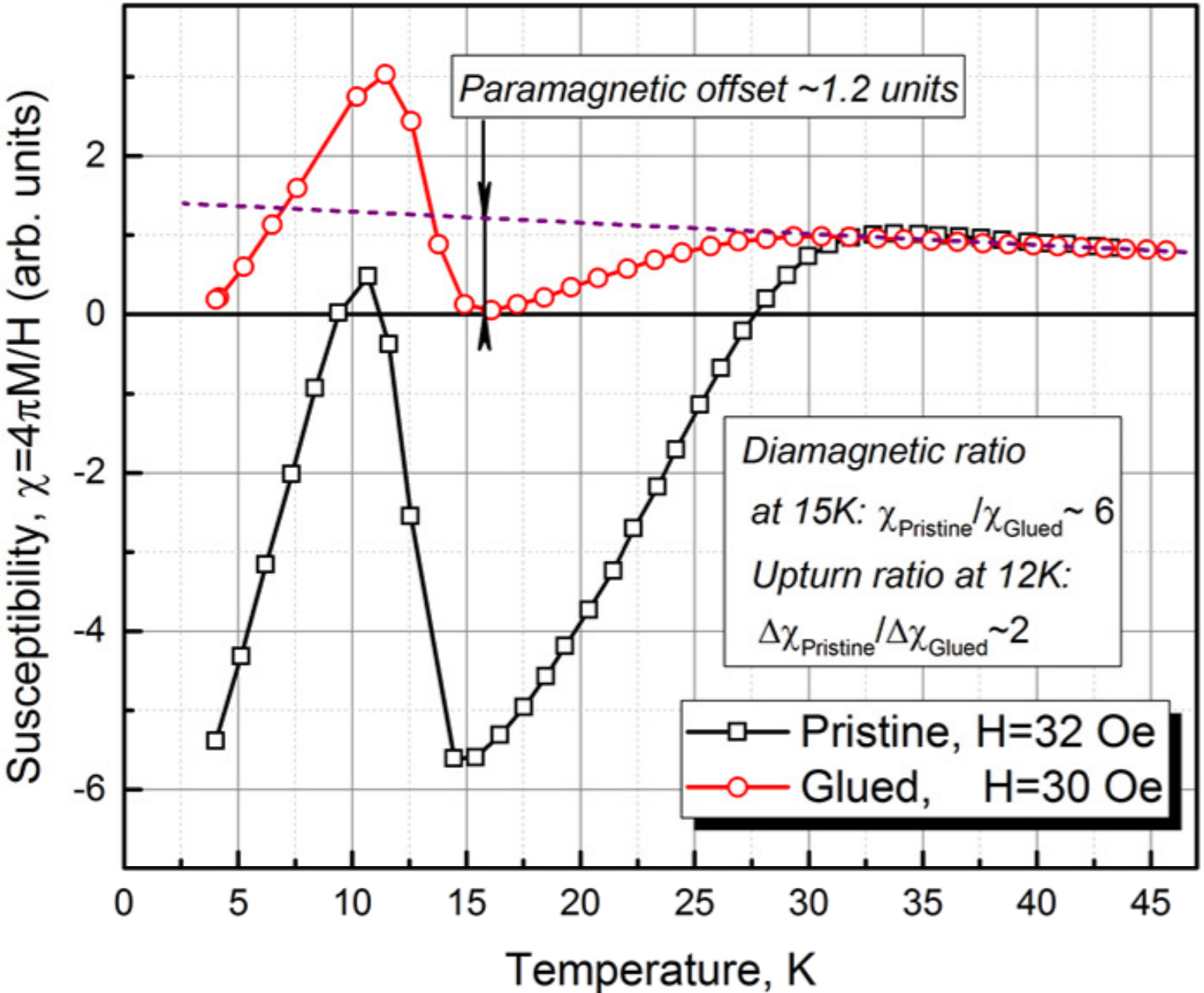}
    \caption{DC susceptibility measurements of pristine and powder-glued sample
in $H\sim $ 30 Oe (similar results were obtained in $H\sim $ 5 Oe).}
    \label{fig7}
\end{figure}

The depth of the first diamagnetic transition (with an onset at 30-35 K),
normalized to the paramagnetic signal, is drastically reduced. That is, for
the same amount of contributing mass of YBa$_{2-x}$Sr$_x$Cu$_{3}$O$_{7-x}$Se$_{x}$, the
diving amplitude of the first diamagnetic signal is much smaller (by a
factor of 6 to 7). This can be explained by the idea that the first
transition is related with the supercurrents flowing on the surface of the
granules as well as the intergranular trajectories (the latter have been
essentially destroyed by milling and subsequent application of the glue). As
a result, the effective volume able to conduct screening superconducting
currents is reduced. At the same time, the upturn amplitude is smaller only
by a factor of two. Importantly, the temperature of upturn also did not
change; most likely, this is because the second superconducting transition
takes place inside of the granules and is not significantly impacted by the
granular size reduction and the granules separation.

Overall, one can conclude that spontaneous currents due to the $\pi -$%
junctions, while being very elegant and possibly applicable to other
systems, are not the cause of the Wohlleben effect in our material. Flux
compression looks more feasible. This model as formulated by Koshelev and
Larkin \cite{KoshelevLarkin}, has been developed in detail by many works
(see, e.g., \cite{Zharkov2001} and Refs. therein). In granular
superconductors, complete screening of applied external field $\mathbf{H}_{ext}$ is
not the only possibility, though it corresponds to the absolute minimum of
Gibbs' energy. In some cases, it may happen that superconductivity first
occurs on the granular surface. Then a flux may be trapped inside of a
granule being confined by the superconducting shell layer. When the
temperature is further reduced, the superconducting layer grows inwards, and
a screening current of the same nature which initially occurred on the
surface layer to screen-off the $\mathbf{H}_{ext}$ rises around the trapped flux. From
the topology it is clear that the direction of this current should be
opposite to that of the surface one. In accordance to (\ref{A1}), this can
revert the sign of magnetization, making it positive.

Using expressions (6) and (13) in the original work  
\cite{KoshelevLarkin}, one can obtain maximum paramagnetic to diamagnetic
susceptibility ratio $\sim 100\%$ for the the disk geometry. This value is
very close to our data in Fig. 7 for the pristine sample.

One can expect the higher$-T_{c}$ phase to be on the surface of individual
granules (with the lower Se content), while lower$-T_{c}$ phase (more rich with Se
content) is inside the granules. This structural order is similar to the
In-Sn$-$composition studied by Chu et al. \cite{Chu2006}. In their study,
the lower$-T_{c}$ superconducting volume was surrounded by a higher$-T_{c}$
phase, and the Wohlleben effect was detected during the
field cooling. From the point of view of the underlying physics, this
structural order facilitates flux trapping and subsequent compression.
Unsurprisingly, the resultant curves (Fig. 2 of Ref. \cite{Chu2006}) resemble
ours shown in Fig. 5\textbf{c} and Fig. 7.

\subsection{Heat capacity anomaly}
The heat capacity measurements and related results (Fig. 6) yield the major
implications of our work, as will be clear from further discussions. The
low-temperature superconducting phase, corresponding to $T_{c2}=12.8$ K,
constitutes the majority. Thus, in our analysis of the heat capacity data,
we will focus on this phase. Excluding the phonon and Schottky contributions
(the details are provided in Methods), we find (Fig. 6\textbf{b})
that a rather good theoretical fit to our experimental data is possible
within the $d-$wave $\alpha -$model (see, \textit{e.g}., \cite%
{Johnston2013,Huang2006}). As follows from this fit, the superconducting
phase content is about $42(wt.\%)$ of the normal phase. This is not very
far from the XRD results (30.6 \textit{at.}\% in Table II correspond to 44.4 \textit{wt.}\%.

Figure 6c,d contains a very unusual result. Namely, in the range $H=0-5$ T,
with the $\textbf{H}-$field increase, the transition temperature $T_{c2}$ moves
towards lower values and the ratio $C_{el}(T)/\gamma T$ equals $1$ at $%
T>T_{c2}$ in accordance with expectations. However, for further increase of
magnetic field (i.e., at $H=7$ T and much more pronouncedly at $H=9$ T), $%
C_{el}(T)/\gamma T$ exhibits a downturn at temperature $T>T_{c2}$ (the
reproducibility of this result has been confirmed on more than one sample;
moreover, to eliminate possibility of data processing artifacts, we checked
visibility of the downturn by comparing the 9 T $C_{p}/T$ raw data with
those in smaller fields: 7 T and below). The field 9T is still below $H_{c2}$%
, as follows from the resistivity curve in Fig. 5\textbf{e}, $T_{c2}\sim 6$
K at $H=90,000$ Oe. At this temperature, the electronic heat capacity has a
local maximum, Fig. 6\textbf{c}, which is still noticeable after the subtraction of
the Schottky part (Fig. 6\textbf{d}). Remarkably, the subtraction of
Schottky contribution does not significantly affect the onset temperature of
the downturn. In absence of the downturn, the curve in case of $H=9$ T (as
well as $H=7$ T) would have had the traditional shape (i.e., similar to
those at smaller values of $H$). 

Thus, we have an anomaly, which is shown in Fig. 6\textbf{c},\textbf{d}, and
is magnified in insets to them. Because of importance of this downturn (see
the next subsection) we also performed measurements at higher fields $H=10$ T, 11 T, 
12 T and 13 T. The results are summarized in Fig. 6\textbf{e} for
more convincive evidence of the effect to be discussed in the next
subsection. 

Prior to proceeding to it, we will mention that the higher $T_{c}$
transition, the $T_{c1}$-phase, is also visible in the
top inset to Fig. 6\textbf{a}. As was mentioned at discussing the
Wohlleben effect, it may be just the crust contribution of the same bulk
Phase 1 with less content of Se. Its field dependence contains another
anomaly: the higher transition temperatures at higher fields. However,
further discussion of this minor phase behavior is beyond the goals of this
report.

\subsection{Quantum criticality}
Magnetic field-induced quantum phase transitions have been reported in
heavy-fermion \cite{Custers2003,Bianchi2003,Sidorov2002}, iron-based \cite%
{Analytis2014,Grinenko2017}, and cuprate \cite%
{Nakajima2007,Valla1999,Keimer2015,Sachdev2010} superconductors, as well as
in metamagnets \cite{Zhou2004,Grigera2004}, antiferromagnets \cite%
{Liu2020,Das2019}, and other systems \cite{Sachdev_2011}. Exploration of
this important quantum-mechanical feature of superconducting state appears
instrumental for deciphering the mechanism of high-temperature
superconductivity.

In the case of quantum criticality, nearly degenerate ground states exist
not only at very low (ideally zero) temperatures, but reveal themselves over
a range of temperatures. Variation of external or internal parameters may
cause border-crossing between the phases on phase diagrams. In the case of
the second order phase transition, the system may possess a set of quantum
critical points (QCPs) at certain values of tuning parameters \cite%
{Keimer2015}. It was recently recognized that the pseudogap (PG) phase of
cuprates ends at QCP \cite{Michon2019}. Moreover, the associated
fluctuations can be involved in $d-$wave pairing and the anomalous
scattering of charge carriers. Heat capacity measurements in the magnetic
field \cite{Michon2019} are instrumental for the observation of features
relevant to quantum criticality. However, for cuprates, the critical fields
are high, and the required values (e.g., 18 T in case of Eu-LSCO and
Nd-LSCO \cite{Michon2019}, and 45 T in case of Tl-2201 \cite%
{Shibauchi2008}) exceed typically available limits of common laboratory
apparatus. In case of Tl-2201, the heavily doped composition is chosen to
reduce the value of $T_{c}$ (and associated destructive magnetic field) from
maximally available $T_{c}=93$ K down to 15 K. In cuprates,
compositional modifications may serve as an alternative to oxygen
overdoping, thus reducing the values of required fields while keeping the
typical features observable.

To understand the relationship of the heat capacity anomaly of the bulk
phase with the quantum criticality, one should take into account that $%
C_{el}(H)/T\equiv \gamma (H)\varpropto n(H)$, where $n$ is the charge
carrier density. If at $H>H_{0}$ (where $H_{0}\simeq 7$ T) $n(T)$ decreases
with the temperature then a non-Fermi liquid behavior takes place: in Fermi
liquids, $n$ (and $\gamma $) does not depend on temperature. This kind of
effect caused by application of strong magnetic fields was reported and
explained by the fluctuating pair field and PG, which causes the downturn
via the border crossing due to the doping \cite%
{Kordyuk2015,Cyr-Choiniere2018,Calegari2015}. In our case, the crossing of
the border is due to the cooling in a high magnetic field. We suggest the
explanation of its mechanism 
which is related with the existence of pair density wave (PDW) \cite%
{Agterberg2020}, and the charge density waves (CDW) as its consequence 
\cite{Choubey2020}. The emergence of CDW requires the doping range $0.08<$ $%
p_{CDW}<0.16$. Within this doping level, in high-enough magnetic fields
(i.e., at $H>H_{0}$), the CDW border crossing occurs as soon as the
temperature drops below a certain value. In pure YBCO, this occurs at $T<20$
K in $H=50$ T \cite{Badoux2016,LeBoeuf2011}. During this process, the
re-arrangement of phases inherent to cuprate superconductors relative to the
QCPs on the field-doping phase diagram takes place.

Appearance of CDW reduces the density of charge carriers. This is associated
with the creation of electron pockets\footnote{%
Interestingly, in addition to the electron pockets on the Fermi arcs,
experimental evidence is obtained in favor of two nearby hole-type pockets 
\cite{Doiron-Leyraud2015}, so that the electronic specific heat is a
combination of three contributions, of which the hole pockets contribute
about $17.3\%$ each to $65.4\%$ of the electron pocket contribution. Unlike
the Hall effect, these electron and hole pocket charge carriers contribute
constructively to the specific heat thus increasing the overall effect of
the magnetic field.} on Fermi arcs, and this mechanism reveals itself in
Hall \cite{Badoux2016,LeBoeuf2011}, Nernst \cite{Doiron-Leyraud2015} and
Seebeck \cite{Laliberte2011,Doiron-Leyraud2015} effects. In particular, the
Hall coefficient $R_{H}$ for $H$ $>H_{0}$, decreases from its zero
value and becomes negative at low-enough temperatures. The increase of the
absolute value of $R_{H}\propto 1/n$ corresponds to the decrease in charge
carrier density $n$ (see \cite{Badoux2016}).

An alternative explanation involves the incommensurate SDW \cite{Haug2010}
rather than CDW. The latter will be prohibited if $p<0.08$. If Se, like O,
provides doping to the CuO conductivity layer, the doping parameter $p$
(i.e., the number of holes per Cu) may be estimated by the universal trend
reported in Ref. \cite{Presland1991}:

\begin{equation}
\frac{T_{c}}{T_{c}^{\max }}=1-82.6(0.16-p)^{2}.  \label{3}
\end{equation}%
Here, $T_{c}^{\max}$ is the value of $T_{c}$ at the optimal doping for our
material. The CDW condition $0.08<p<0.16$ \cite{Hucker2014,Blanco2014}
infers a range of possible values for $T_{c}^{\max }$: $12.8$ K $%
<T_{c}^{\max }<27.2$ K. The inhomogeneity-caused broad range of $T_{c}$
values visible in Fig.~6b allows the emergence of quantum criticality at
15~K which is in accordance with the downturn temperature $\sim $15~K of 9 T
and higher field-caused QCP in Fig. 6\textbf{c}-\textbf{e}. As known for HTSC cuprates 
\cite{Kordyuk2015,Cyr-Choiniere2018,Calegari2015}, at $T=0$, $p_{PG}=0.19$.
Corresponding value of $T_{c}^{max}$ from (\ref{3}) is then $\sim $14~K,
which, within the experimental inaccuracy, is close to the declared 15 K.
The fact that the QCP observational temperature is dramatically higher than $%
T=0$ in both competing cases is not surprising in view of the
well-established ``fanning out" effect \cite{Coleman2005}
reported previously \cite{Keimer2015,Badoux2016} in HTSC cuprates.

For pure single-crystal measurements, involvement of quantum criticalities
yields peculiarities in resistivity $\rho $. For example, a drastic upturn
takes place at PG when superconductivity is fully suppressed \cite%
{Cyr-Choiniere2018} by $H>H_{c2}$: the upturn starts gradually much below
the pseudogap opening temperature $T_{\rho }$. Also, two competing
mechanisms (reduction of inelastic scattering and reduction of charge
carrier density) may coexist, so the resultant behavior of $\rho (T)$ may be
both upturn and downturn. In our case, $H_{max}=9$ T $< H_{c2}$. 
Additionally, in our case, we have the second, higher-$T_{c}$ phase which
can mask-off any gradual changes in $\rho $; moreover, as follows from the
inset in Fig. 5\textbf{e}, granular crusts are causing semiconductor-type temperature
behavior of $\rho (T)$ above $T_{c1}$ thus obscuring involvement of
arguments related with the reported behavior of resistivity. Having single
crystals of this composition where the value of $p$ depends on Se
concentration $y$ would further facilitate research in these cuprates. In
particular, one can explore oscillations of the Seebeck and Nernst
coefficient (see Fig 2 in \cite{Doiron-Leyraud2015}) and other effects
caused by quantum criticality in much more accessible magnetic fields. In
the course of our research the doping parameter was kept constant;
principally, additional opportunities may come from variation of Se
concentration as a QCP tuning parameter.

In general, simultaneous doping of mother substance YBa$_{2}$Cu$_{3}$O$_7$ 
by Sr and Se provides more flexibility in modifying this
remarkable system than mono-element substitution. The price to pay for this
flexibility is multiphase nature of the sample. However, after this study,
one can try to target more precisely the composition YBa$_{1.4}$Sr$_{0.6}$Cu$_{3}$O$_{6}$Se$_{0.51}$ 
 with the adjusted initial stoichiometry - a fair task for future experimental 
 research. Also, explaining why its properties are drastically different 
 from the isomorphic YBa$_2$Cu$_3$O$_{6.51}$ would be very challenging 
 for the theory.

\section{Conclusions}

In summary, we obtained striking anomalies delivered by simultaneous double
substitution of Ba by Sr and O by Se in a pristine YBa$_{2}$Cu$_{3}$O$_{7}$
superconductor. Simultaneous application of EDX and EBSD proved that Se
enters into the crystal lattice of the superconducting phase of
polycrystalline heterogeneous material. Two superconducting transitions are
revealed. One phase, with $T_{c1}\approx 35$ K, and the second one, with
lower $T_{c2}\approx 13$ K. The latter phase, 
YBa$_{1.4}$Sr$_{0.6}$Cu$_{3}$O$_{6}$Se$_{0.51}$ 
constitutes the majority of
the superconductor. The second transition, most likely is caused by the granular surface of the same 
substance with less amount of Se on the surface of granules. Heat capacity
explorations proved the $d-$wave nature of the main phase, as well as the
existence of electronic Schottky anomaly in it. Susceptibility measurements
revealed magnetic clusters below the Curie temperature $\approx 4$~K. These
magnetic moments, related with Se ligands of Cu, are reported for the first
time. In view of provided XRD information on crystallographic positions of
elements in superconducting phase, fruitful theoretical exploration should
become possible, as it happened with As vacancies in LaO$_{0.9}$F$_{0.1}$FeAs%
$_{1-\delta }$ ($\delta \approx 0.06$) \cite{Grinenko2011}. The most
important finding in our report is related with the anomaly in the heat
capacity data in moderately strong ($7-9$~T) magnetic fields. Various
scenarios related with the competing PDW and PG QC border crossing can
explain the obtained effect. Further experimental work and theoretical
analysis on coexistence of PDW/PG and superconductivity will be easier in
YBa$_{2-x}$Sr$_x$Cu$_3$O$_{7-x}$Se$_x$ than in pure YBCO
due to the less stringent magnetic field requirement, preferably on
single-crystalline or single-phase samples, and could reveal more understanding of
interrelationship between the physical effects reported here and the
mechanism of superconductivity in cuprates.

\section{Methods}

\subsection{{Materials preparation}}
Polycrystalline samples were prepared via standard solid-state synthesis
routes. Powders of \textrm{Y}$_{2}$\textrm{O}$_{3}$, \textrm{SrCO}$_{3}$, 
\textrm{BaCO}$_{3}$, \textrm{CuO} and \textrm{SrSe} were mixed in
stoichiometric proportions to form, after calcination, initial composition 
YBa$_{2-x}$Sr$_x$Cu$_3$O$_{7-x}$Se$_x$ ($x=0$, 0.5, 0.75, 1). After thorough mixing
(hand mixing in sapphire mortar for $40-50$ \textrm{min} per $400$ \textrm{mg%
} of total constituents, or, alternatively, combination of initial $10$ 
\textrm{min} of hand mixing with $4\times 5$ \textrm{min} of mechanical mill
mixing (in plastic vials with agate balls), the mixture was calcined at $%
900^{0}$\textrm{C} for $100$ \textrm{min} in KSL-1100 furnace in air, then
re-ground, milled again, and pelletized for the second thermal treatment,
which was done at $950^{0}$\textrm{C} for $30$ \textrm{min}. It was then
continued at $650^{0}$\textrm{C} for \textrm{80} \textrm{min}. Though the
elemental composition of synthesized samples may be different from the
initial targeted stoichiometry, we still call the resultant composition as 
YBa$_{2-x}$Sr$_x$Cu$_3$O$_{7-x}$Se$_x$. Several pellets
were prepared, with diameter $4$ \textrm{mm}, height $1.5$ \textrm{mm} for
resistivity and magnetic measurements, and with diameter $1.5$ \textrm{mm}
and height $2$ \textrm{mm} for heat capacity measurements.
The most interesting features were discovered in case of the initial 
stoichiometry with $x=1$, so only this case is being discussed in this report.

\subsection{{Compositional Analysis}}
For quantitative EDX analysis, ion milling was performed. The sample surface
was polished so as to avoid the undesired self-absorption which affects
proper outcome for light elements (O and Se in particular). For this task,
the fine-ground powder was suspended in isopropyl alcohol, and dropped into
a gap between a stainless steel holder and aluminum foil touching it as
shown in Fig. 8\textbf{a}. 

\begin{figure}
    \centering
    \includegraphics[width=\linewidth]{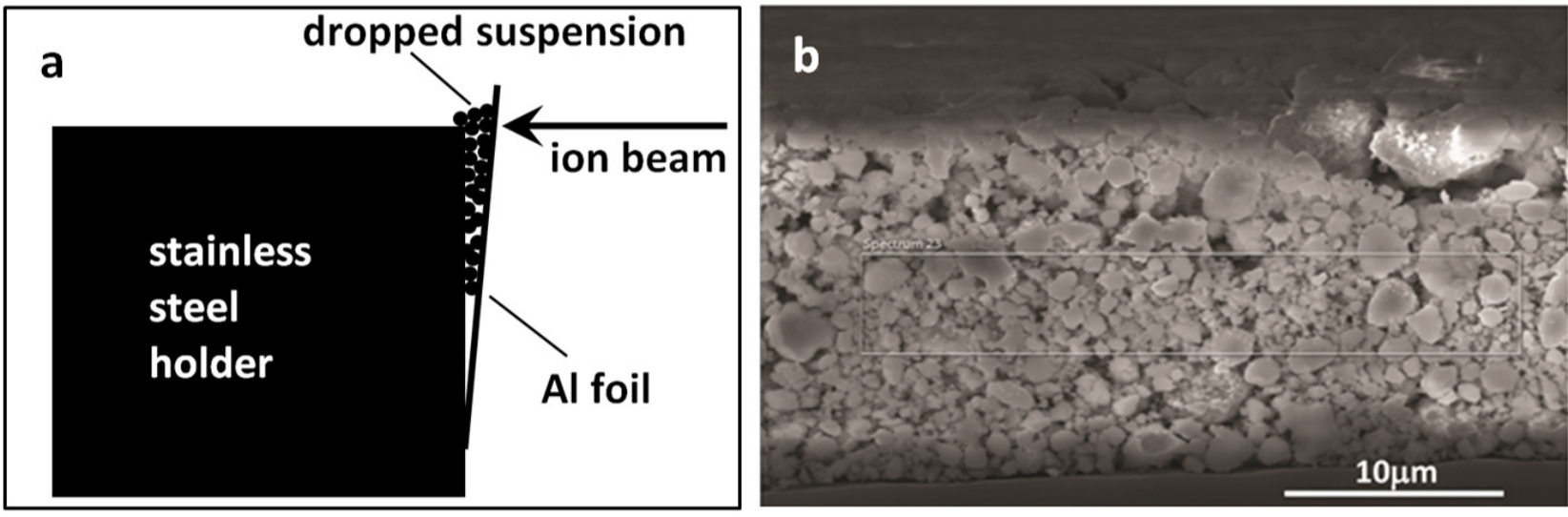}
    \caption{Polishing the powder via ion-milling. (\textbf{a})
Arrangement for cutting by ion knife. (\textbf{b}) Surface of powder
prepared for EDX after the ion cutting.}
    \label{fig8}
\end{figure}

The cutting by ion knife was arranged in Hitachi Broad Ion Milling System,
IM4000Plus, which cut the structure horizontally, providing a polished
surface for microanalysis (Fig. 8\textbf{b}). The SEM EDX data were
acquired from a large surface area, and the revealed composition closely
resembled the initial chosen stoichiometry of the specimen: the data
reflected in Table 1 (last line) confirm that a rather significant amount of
Se remains in the material.

A Hitachi SU5000 variable pressure field emission scanning electron
microscope (VP-FESEM) with Oxford Instruments EDX and EBSD under AZtec
software was used for simultaneous EDX/EBSD analysis with electron beam
focal diameter $\sim 1-10$ nm which is suitable for the proposed task since
it is comparable with the size of the lattice cell. During this stage of
exploration, no ion polishing was applied. Instead, in order to avoid
self-absorption, we changed the incidence angle of the electron beam until a
Kikuchi pattern appeared (Fig. 3\textbf{b}) for identifying the crystalline
structure (Fig. 3\textbf{c}). We then determined the chemical composition
via EDX (Fig. 3\textbf{d}) at the same region of interest.

\subsection{{X-ray diffraction studies}}
Diffraction experiments were performed with polycristalline samples in
accordance with the data given in Table 4.

\begin{table}
\centering
\caption{Parameters of the powder diffraction experiment}
\begin{tabular}{|l|l|}  \hline
Diffractometer & Rigaku MiniFlex 600 \\  \hline
Radiation & Cu-K$_{\alpha }$ \\  \hline
Wavelength, \AA  & 1.540593 \\  \hline
Detector name & D/teX Ultra2 \\  \hline
Detector type\qquad & point \\  \hline
Scan range, deg & 10.0 - 110.0 \\  \hline
Scan step, deg & 0.005 \\  \hline
Scan speed, deg/ min & 1.0 \\  \hline
Incident slit, type, deg & Soller slit, 5.0 \\  \hline
Receiving slit, type, deg & Soller slit, 5.0 \\  
\hline
\end{tabular}
\end{table}

Qualitative phase analysis was performed with the code HighScore Plus 3.0e 
\cite{Degen2014}. Heterophase nature of the sample was revealed and for the
explanation of all detected peaks, five phases have been involved.

Refinement of structural models of these phases was performed using Jana2006
software \cite{Petricek2014}. Complex heterophase content of sample yielded
strong correlations between the refining parameters of the models. For the
reduction of the influence of these correlations on the results of our
analysis we used various methods, the main of which was cyclic combination
of Rietveld's approach \cite{Rietveld1967,Rietveld1969} and Le Bail's approach 
\cite{Lebail1988} which significantly increased the stability of the
refinement and much improved the fit. The complexity of the problem
arose from the small amount of Se dopant which cannot provide strong
response at the scattering of X-rays. Nevertheless, because of the achieved
high level of relative precision, we were able to detect the difference between the
three cases shown in Table 5.

\begin{landscape}
\begin{table*}
\centering
\caption{Refinement comparison for three models of Se location in the structure YBa$_{1.4}$Sr$_{0.6}$Cu$_{3}$O$_{6}$Se$_{0.51}$}
\resizebox{\linewidth}{!}{
\begin{threeparttable}
\begin{tabular}{|l|l|l|l|l|l|} \hline
{No}  &   {Model}   &   {\textit{R}1, \%}    &   {\textit{w}, \%}    & {$\mathbf{\Delta\rho _{\mathrm{max}}}$, e/\AA$^3$}  &  {$\mathbf{\Delta\rho _{\mathrm{min}}}$, e/\AA$^3$}     \\ \hline
1 \cite{Licci1998} & Additional O positions: O4($x$,0.5,0) and O5(0.5,0,0); Se is absent in the structure\tnote{a} &   0.98  &   0.90  &   +0.30   &   -0.48   \\ \hline
2 \cite{Yakinci2013}   &   Se partially replaces Y\tnote{b} &   1.09    &   0.93    &   +0.36   &   -0.54   \\\hline
3 Our model &   Se is localized at positions Se4($x$,0.5,0) and Se5(0.5,0,0)\tnote{c}  &   0.87    &   0.73    &   +0.17   &   -0.30     \\ 
\hline
\end{tabular}
\begin{tablenotes}
     \item[a] At full matrix refinement, because of correlational shift, the $x$-coordinate of position O4 is unrealistic.
     \item[b] Such heterovalent substitution looks improbably; we considered this opportunity for the test of literature data. It worsens refinement parameters; the results are unrealistic.
     \item[c] In this model, the full matrix refinement is possible; moreover, with the simultaneous refinement of populations and ADP for incompletely occupied or shifted positions.
   \end{tablenotes}
\end{threeparttable}}
\end{table*}
\end{landscape}

These model results were compared against each other using the criterion of
refinement stability and factors \textit{R}1 and \textit{wR}. It turned out that for the
localization of dopant atoms, the extrema of difference Fourier synthesis, 
$\Delta \rho _{max}/%
\Delta \rho _{min}$ are more informative. This criterion is usually considered as the second
main one for the refinement of structural models. The main superconducting
phase was determined in Model 3 with significant reduction of $R$-factors for
the same number of Bragg reflections (152) and refinement parameters (18) as
with Models 1 and 2. The highest residual peaks of difference
electronic density are localized near heavy cations. The details of the
refinement are given in Table 6.

\begin{table}
\centering
\caption{Details of structural refinement for 
YBa$_{1.4}$Sr$_{0.6}$Cu$_{3}$O$_{6}$Se$_{0.51}$ crystaline phase} 
\resizebox{\linewidth}{!}{
\begin{tabular}{|l | l|} \hline
Temperature (K) & 294 \\ \hline
Crystal system & Orthorhombic \\  \hline
space group, $Z$ & $Pmmm$, 1 \\  \hline
$a, b, c$ (\AA ) & 3.84359(4), 3.83295(6), 11.47722(17) \\  \hline
$V$ (\AA $^{3}$) & 169.092(4) \\  \hline
$Dx$ (g/cm$^{3}$) & 6.8015 \\  \hline
\textmu (mm$^{-1}$) & 83.432 \\  \hline
\textit{Profile fit} &  \\  \hline
Profile function & Pseudo-Voigt \\  \hline
No. of measured points & 20001 \\  \hline
$R_p$, \%\qquad & 1.32 \\  \hline
$wR_p$, \% & 1.92 \\  \hline
\textit{GOF} (S) & 3.14 \\  \hline
\textit{Structure model fit} &  \\  \hline
Refinement based on \textit{F} with a weighting scheme & 1/$\sigma ^{2}$(\textit{F}) \\  \hline
No of Bragg reflections, parameters\qquad & 152, 18 \\  \hline
$R1$, \%, \qquad & 0.87 \\  \hline
$wR$, \%\qquad & 0.73 \\  \hline
$\Delta \rho _{\max }/\Delta \rho _{\min }$, e/\AA $^{3}$\qquad & +0.17 / 
-0.30 \\  \hline
Software\qquad & HighScore Plus, Jana2006 \\
\hline
\end{tabular}}%
\end{table}

\subsection{{Processing of heat capacity data}}
Assuming that (at low temperatures) the heat capacity in our samples has
three major contributions, we can write%
\begin{equation}
C(T)=C_{el}(T)+C_{ph}(T)+C_{Sch}(T),  \label{4}
\end{equation}%
where $C_{el}(T)$ stands for the electronic contribution containing both the
normal part $\gamma T$\ and the superconductor part, $C_{ph}(T)=\beta T^{3}$
characterizes the phonon contribution, and the last term is the Schottky
anomaly, which, in case of the simplest two-level gap $\Delta _{tl}$ can be
represented as \cite{Xie2012} 
\begin{equation}
C_{Sch}(T)=R\left( \frac{\Delta _{tl}}{T}\right) ^{2}\frac{\exp \left(
\Delta _{tl}/T\right) }{\left[ 1+\exp \left( \Delta _{tl}/T\right) \right]
^{2}}.  \label{5}
\end{equation}

At the first stage, the phonon contribution was subtracted from the
experimental data\footnote{It is worth to mention that our heterophase substance contains ``parasitic'' contributions to the heat capacity of the superconducting Phase 1. At $T=15$ K, the $C/T$ of major phases $2-4$ are (in mJ/(K$^{2}$ g-at) units): $\sim 1$ for BaSeO$_{4}$ \cite{BaSeO4}, $\sim 0.52$ for Y$_{2}$Cu$_{2}$O$_{5}$  \cite{Matskevich-arxiv}, and  $\sim 20$ for Y$_{2}$BaCuO$_{5}$ \cite{Eckert1988}. These exceed the contribution of the superconducting phase (Fig. 5\textbf{a}). However, their contribution is mainly phononic in origin, and subtraction of $\beta T^{2}$ from $C/T$ extinguishes it as well as the contribution of the minor phase BaCu(SeO$_{3}$)$_{2}$.}. 
For that task, $C(T)/T$ was plotted as a function of $%
T^{2}$. Since the Schottky contribution (\ref{5}) is negligible
(exponentially small) at $T=0$, the inclination of $C(T)/T$\ near the origin
of coordinates provides the value of $\beta $ (in our case, $\beta \approx
0.04$ (Fig. 6). Also, the value of $C(T)/T$\ \ at $T=0$ equals $\gamma $.
After the subtraction of phonon contribution, the deviation from typical
electronic contribution of superconducting state becomes evident (cf. Fig.
6\textbf{c}-{e}).

The characteristic superconducting behavior is restored by the subtraction
of the Schottky contribution (\ref{5}). After these operations are
performed, one can make the second iteration, which involves computer
coding. In the first step in this code, the Schottky curve is subtracted
from the experimental data in accordance to (\ref{4}) while allowing the
opportunity to vary the gap $\Delta _{tl}$ and the prefactor $R$ around the
values obtained at the initial stage. Then, the code computes the
corresponding values of $\beta $\ and $\gamma $. These values are not too
far from the ones determined at the initial stage, but\ performing the
iteration is useful for ensuring better data processing for further
analysis. The curves in Fig. 6\textbf{d} were obtained via this procedure
from Fig. 6\textbf{c}.

For the theoretical modeling (Fig. 6\textbf{b}), we used the standard
expression for heat capacity \cite{Huang2006}, which, in dimensionless form,
can be represented as:

\begin{equation}
\begin{split}
\frac{C_{el}(t)}{T_{c}}=\frac{1}{4t^{2}}&\int_{0}^{2\pi }d\varphi
\int_{0}^{\pi }d\theta \sin \theta \\
&\int_{-\infty }^{\infty }dx\frac{%
x^{2}+\delta (t)^{2}-t\delta (t)\left[ d\delta (t)/dt\right] }{\cosh ^{2}%
\left[ \sqrt{x^{2}+\delta (t)^{2}}/\left( 2t\right) \right] },
\end{split}
\label{6}
\end{equation}%
where: $T/T_{c}\equiv t;$\ $\xi /T_{c}\equiv x;\Delta (T)/T_{c}\equiv \delta
(t);$ $\varepsilon /T_{c}\equiv \sqrt{\xi ^{2}+\Delta ^{2}}/T_{c}=\sqrt{%
x^{2}+\delta ^{2}}.$ In (\ref{6}), one should substitute the value of the
order parameter (the gap), which has the form $\Delta =\Delta _{0}$ for
isotropic $s-$wave and $\Delta =\Delta _{0}\cos n\phi $ for line nodes ($n=2$
for $d-$wave $\alpha -$model) \cite{Johnston2013,Lin2011}. The temperature
dependence of\ $\Delta (T)$\ was defined via the analytical expression \cite%
{Carless1983} for BCS solution:%
\begin{eqnarray}
\frac{\Delta }{\Delta (0)} &=&\sqrt{1-t}(0.9663+0.7733t),\text{if }t>0.33 
\nonumber \\
&=&1\text{ otherwise. }  \label{7}
\end{eqnarray}

Computation of $C_{el}(T)/T$\ based on (\ref{6}) and (\ref{7}) was
performed numerically both for the $s-$wave and the $d-$wave cases. The best
result is shown in the Fig. 3\textbf{b}. It corresponds to the Schottky
parameters $\Delta _{tl}=10.5\pm 0.1$ K, $R=0.31\pm 0.01$. This value of $%
\Delta _{tl}$ is close the Schottky gaps in YBCO materials \cite{Collocott88}. 
This matching justifies application of the simple relation (\ref{5})
instead of more sophisticated expression \cite{Mu2007}: 
\begin{equation}
C_{Sch}=R\frac{dE_{Sch}}{dT},  \label{8}
\end{equation}%
where%
\begin{equation}
E_{Sch}=\sum E_{i}\exp (-E_{i}/T)/\sum \exp (-E_{i}/T),  \label{9}
\end{equation}%
and $E_{i}=g\mu _{B}M_{J}H_{eff}$ where $M_{J}=\{-S,-S+1,...S-1,S\}_{i}$ , $%
H_{eff}=\sqrt{H_{0}^{2}+H_{ext}^{2}}$ where $H_{0}$ and $H_{ext}$ are the
internal and external field values. In the case of double levels ($S=1/2$), (%
\ref{8}) transforms into (\ref{5}) and the value of $\Delta _{tl}$ is
expressed by the external and the internal fields $H_{ext}$ and $H_{0}$. In
the case of $S=5/2$ (which corresponds to the $2+$ states of copper ions),
we obtained that the same level of matching as in Fig. 6\textbf{b} is
possible at somewhat lower values of $H_{0}$. Whichever case is more
relevant to our material requires further analysis. Also, the physical cause
of the Schottky anomaly in our case remains uncertain: in addition to
scattering on magnetic clusters, the scattering on minor amount of rare
earth impurities \cite{Collocott88} is an alternative. However, the details
of Schottky anomaly are not significant for our main results, and we will
not dwell more on this topic.

\section{Data availability}

The data that support the findings of this study are available from the corresponding author upon reasonable request.

\section{Code availability}

Full code that support the findings of this study are available from the corresponding author upon reasonable request.

\section{Acknowledgments}
A.D. acknowledges support by the Ministry of Science and Higher Education of
the Russian Federation within the State assignment FSRC "Crystallography and
Photonics" RAS in part of X-rays diffraction study. The work of the Chapman
U. research team is supported by the US Office of Naval Research Grants No.
N00014-21-1-2879 and N00014-20-1-2442. We are grateful to E. Vinogradova for assistance.

\section{Author contributions} 
A.G., V.N. , V.G. and S.N. designed research; A.G., V.G., V.N., J.K., A.M., J.C., I.P, R.D., S.T., T.H. and S.C. performed research, A.G., V.N., V.G., A.D., R.D., S.T. and S.C. analyzed data; and  A.G., V.G., A.D., R.D., S.T. and S.C. wrote the paper.

\appendix


\bibliographystyle{elsarticle-num} 

\begin{thebibliography}{10}
\expandafter\ifx\csname url\endcsname\relax
  \def\url#1{\texttt{#1}}\fi
\expandafter\ifx\csname urlprefix\endcsname\relax\def\urlprefix{URL }\fi
\expandafter\ifx\csname href\endcsname\relax
  \def\href#1#2{#2} \def\path#1{#1}\fi

\bibitem{Onnes1910}
H.~{Kamerlingh Onnes}, {Further experiments with liquid helium. C. On the
  change of electric resistance of pure metals at very low temperatures etc.
  IV. The resistance of pure mercury at helium temperatures}, Koninklijke
  Nederlandse Akademie van Wetenschappen Proceedings Series B Physical Sciences
  13 (1910) 1274--1276.

\bibitem{matriconBook}
J.~Matricon, G.~Waysand, The Cold Wars: A History of Superconductivity, The
  Cold Wars: A History of Superconductivity, Rutgers University Press, 2003.

\bibitem{Bednorz86}
J.~G. Bednorz, K.~A. M{\~A}{\textonequarter}ller, {Possible high Tc
  superconductivity in the Ba--La--Cu--O system}, Zeitschrift f\"ur Physik B
  Condensed Matter 64~(2) (1986) 189--193.
\newblock \href {https://doi.org/10.1007/BF01303701}
  {\path{doi:10.1007/BF01303701}}.

\bibitem{Wu87}
M.~K. Wu, J.~R. Ashburn, C.~J. Torng, P.~H. Hor, R.~L. Meng, L.~Gao, Z.~J.
  Huang, Y.~Q. Wang, C.~W. Chu, {Superconductivity at 93 K in a new mixed-phase
  Y-Ba-Cu-O compound system at ambient pressure}, Phys. Rev. Lett. 58 (1987)
  908--910.
\newblock \href {https://doi.org/10.1103/PhysRevLett.58.908}
  {\path{doi:10.1103/PhysRevLett.58.908}}.

\bibitem{Sheng88}
Z.~Z. Sheng, A.~M. Hermann, {Bulk superconductivity at 120 K in the
  Tl--Ca/Ba--Cu--O system}, Nature 332~(6160) (1988) 138--139.
\newblock \href {https://doi.org/10.1038/332138a0}
  {\path{doi:10.1038/332138a0}}.

\bibitem{Schilling93}
A.~Schilling, M.~Cantoni, J.~D. Guo, H.~R. Ott, {Superconductivity above 130 K
  in the Hg--Ba--Ca--Cu--O system}, Nature 363~(6424) (1993) 56--58.
\newblock \href {https://doi.org/10.1038/363056a0}
  {\path{doi:10.1038/363056a0}}.

\bibitem{Maeda88}
H.~Maeda, Y.~Tanaka, M.~Fukutomi, T.~Asano, {A New High-Tc Oxide Superconductor
  without a Rare Earth Element}, Japanese Journal of Applied Physics 27~(2A)
  (1988) L209.
\newblock \href {https://doi.org/10.1143/JJAP.27.L209}
  {\path{doi:10.1143/JJAP.27.L209}}.

\bibitem{Yee2014}
C.-H. Yee, G.~Kotliar, {Tuning the charge-transfer energy in hole-doped
  cuprates}, Phys. Rev. B 89 (2014) 094517.

\bibitem{Yee2015}
C.-H. Yee, T.~Birol, G.~Kotliar, {Guided design of copper oxysulfide
  superconductors}, {EPL} (Europhysics Letters) 111~(1) (2015) 17002.
\newblock \href {https://doi.org/10.1209/0295-5075/111/17002}
  {\path{doi:10.1209/0295-5075/111/17002}}.

\bibitem{Palhan1988}
L.~Palhan, A.~Brokman, I.~Felner, M.~Brettschneider, Y.~Yacoby, M.~Weger, {Does
  sulfur replace oxygen in the superconducting YBa$_2$Cu$_3$O$_6$S phase?},
  Solid State Communications 68~(3) (1988) 313 -- 317.

\bibitem{Cloots1991}
R.~Cloots, A.~Rulmont, P.~Godelaine, C.~Hannay, H.~Vanderschueren, M.~Ausloos,
  {Sulphur substitution for oxygen in YBa$_2$Cu$_3$O$_7$ ceramics}, Solid State
  Communications 79~(7) (1991) 615 -- 619.

\bibitem{Cooke1999}
S.~Cooke, J.~Allison, R.~Woods, {High-temperature resistivity measurements of
  YBa$_2$Cu$_3$O$_{7-\delta}$ doped with sulphur}, Solid State Communications
  112~(4) (1999) 229 -- 233.

\bibitem{Gagnon1989}
R.~Gagnon, P.~Fournier, M.~Aubin, A.~H. O'Reilly, J.~E. Greedan,
  {Superconductivity in sulfur-containing $R$-Ba-Cu-O compounds}, Phys. Rev. B
  39 (1989) 11498--11502.
\newblock \href {https://doi.org/10.1103/PhysRevB.39.11498}
  {\path{doi:10.1103/PhysRevB.39.11498}}.

\bibitem{Kambe1988}
S.~Kambe, M.~Kawai, {Effect of S, Se and Te Addition on the Superconductive
  Properties of Ba$_2$YCu$_3$O$_{7-y}$}, Japanese Journal of Applied Physics
  27~(Part 2, No. 12) (1988) L2342--L2344.
\newblock \href {https://doi.org/10.1143/jjap.27.l2342}
  {\path{doi:10.1143/jjap.27.l2342}}.

\bibitem{Yakinci2013}
Z.~D. Yakinci, D.~M. Gokhfeld, E.~Altin, F.~Kurt, S.~Altin, S.~Demirel, M.~A.
  Aksan, M.~E. Yakinci, {Jc enhancement and flux pinning of Se substituted YBCO
  compound}, Journal of Materials Science: Materials in Electronics 24~(12)
  (2013) 4790--4797.
\newblock \href {https://doi.org/10.1007/s10854-013-1476-8}
  {\path{doi:10.1007/s10854-013-1476-8}}.

\bibitem{Slebarski1990}
A.~Slebarski, A.~Chelkowski, J.~Jelonek, A.~Kasprzyk, {Effect of substitution
  of O by Se on the superconductivity of YBa$_2$Cu$_3$O$_7$}, Solid State
  Communications 73~(7) (1990) 515 -- 517.

\bibitem{Felner1987}
I.~Felner, I.~Nowik, Y.~Yeshurun, {Effects of substitution of O by S and Cu by
  Fe on superconductivity in
  Y${\mathrm{Ba}}_{2}$${\mathrm{Cu}}_{3}$${\mathrm{O}}_{7}$}, Phys. Rev. B 36
  (1987) 3923--3925.

\bibitem{Felner1988}
I.~Felner, B.~Barbara, {Effect of sulfur on the superconductivity of
  R${\mathrm{Ba}}_{2}$${\mathrm{Cu}}_{3}$${\mathrm{O}}_{7}$}, Phys. Rev. B 37
  (1988) 5820--5823.

\bibitem{Shibauchi2008}
T.~Shibauchi, L.~Krusin-Elbaum, M.~Hasegawa, Y.~Kasahara, R.~Okazaki,
  Y.~Matsuda, {Field-induced quantum critical route to a Fermi liquid in
  high-temperature superconductors}, Proceedings of the National Academy of
  Sciences 105~(20) (2008) 7120--7123.
\newblock \href {https://doi.org/10.1073/pnas.0712292105}
  {\path{doi:10.1073/pnas.0712292105}}.

\bibitem{Michon2019}
B.~Michon, C.~Girod, S.~Badoux, J.~Ka{\v{c}}mar{\v{c}}{\'i}k, Q.~Ma,
  M.~Dragomir, H.~A. Dabkowska, B.~D. Gaulin, J.-S. Zhou, S.~Pyon, T.~Takayama,
  H.~Takagi, S.~Verret, N.~Doiron-Leyraud, C.~Marcenat, L.~Taillefer, T.~Klein,
  {Thermodynamic signatures of quantum criticality in cuprate superconductors},
  Nature 567~(7747) (2019) 218--222.
\newblock \href {https://doi.org/10.1038/s41586-019-0932-x}
  {\path{doi:10.1038/s41586-019-0932-x}}.

\bibitem{Licci1998}
F.~Licci, A.~Gauzzi, M.~Marezio, G.~P. Radaelli, R.~Masini,
  C.~Chaillout-Bougerol, {Structural and electronic effects of Sr substitution
  for Ba in
  $\mathrm{Y}({\mathrm{Ba}}_{1\ensuremath{-}x}{\mathrm{Sr}}_{x}{)}_{2}{\mathrm{Cu}}_{3}{\mathrm{O}}_{w}$
  at varying w}, Phys. Rev. B 58 (1998) 15208--15217.
\newblock \href {https://doi.org/10.1103/PhysRevB.58.15208}
  {\path{doi:10.1103/PhysRevB.58.15208}}.

\bibitem{Pistorius1962}
C.~Pistorius, M.~Pistorius, Lattice constants and thermal-expansion properties
  of the chromates and selenates of lead, strontium and barium, Zeitschrift
  f\"ur Kristallographie 117~(4) (1962) 259--272.

\bibitem{Adams1992}
R.~D. Adams, J.~A. Estrada, T.~Datta, {Crystal structure analysis of
  Y$_2$Cu$_2$O$_5$}, Journal of Superconductivity 5~(1) (1992) 33--38.

\bibitem{Aride1989}
J.~Aride, S.~Flandrois, M.~Taibi, A.~Boukhari, M.~Drillon, J.~Soubeyroux, {New
  investigation on magnetic and neutron diffraction properties of
  Y$_2$Cu$_2$O$_5$ and related oxides}, Solid State Communications 72~(5)
  (1989) 459--463.

\bibitem{Hsu1996}
R.~Hsu, E.~N. Maslen, N.~Ishizawa, {A synchrotron X-ray study of the electron
  density in Y${\sb 2}$BaCuO${\sb 5}$}, Acta Crystallographica Section B 52~(4)
  (1996) 569--575.
\newblock \href {https://doi.org/10.1107/S0108768196000250}
  {\path{doi:10.1107/S0108768196000250}}.

\bibitem{Effenberger1987}
H.~Effenberger, {Three modifications of BaCu(SeO$_3$)$_2$ and the compound
  SrCu(SeO$_3$)$_2$: Preparation and crystal structure determination}, Journal
  of Solid State Chemistry 70~(2) (1987) 303 -- 312.
\newblock \href {https://doi.org/https://doi.org/10.1016/0022-4596(87)90069-7}
  {\path{doi:https://doi.org/10.1016/0022-4596(87)90069-7}}.

\bibitem{Toby2006}
B.~H. Toby, {R factors in Rietveld analysis: How good is good enough?}, Powder
  Diffraction 21~(1) (2006) 67–70.
\newblock \href {https://doi.org/10.1154/1.2179804}
  {\path{doi:10.1154/1.2179804}}.

\bibitem{Gulian2015}
A.~Gulian, V.~Nikoghosyan, J.~Tollaksen, V.~Vardanyan, A.~Kuzanyan,
  {Current-biased Transition-edge Sensors Based on Re-entrant Superconductors},
  Physics Procedia 67 (2015) 834 -- 839, proceedings of the 25th International
  Cryogenic Engineering Conference and International Cryogenic Materials
  Conference 2014.

\bibitem{Levy1968}
R.~A. Levy, {Principles of Solid State Physics}, Academic Press, New York,
  1968.

\bibitem{Grinenko2011}
V.~Grinenko, K.~Kikoin, S.-L. Drechsler, G.~Fuchs, K.~Nenkov, S.~Wurmehl,
  F.~Hammerath, G.~Lang, H.-J. Grafe, B.~Holzapfel, J.~van~den Brink,
  B.~B\"uchner, L.~Schultz, {As vacancies, local moments, and Pauli limiting in
  LaFeAs${}_{1\ensuremath{-}\ensuremath{\delta}}{\mathrm{O}}_{0.9}{\mathrm{F}}_{0.1}$
  superconductors}, Phys. Rev. B 84 (2011) 134516.
\newblock \href {https://doi.org/10.1103/PhysRevB.84.134516}
  {\path{doi:10.1103/PhysRevB.84.134516}}.

\bibitem{Yeshurun1987}
Y.~Yeshurun, I.~Felner, H.~Sompolinsky, {Magnetic properties of a
  high-${T}_{c}$ superconductor
  ${\mathrm{YBa}}_{2}$${\mathrm{Cu}}_{3}$${\mathrm{O}}_{7}$: Reentry-like
  features, paramagnetism, and glassy behavior}, Phys. Rev. B 36 (1987)
  840--842.
\newblock \href {https://doi.org/10.1103/PhysRevB.36.840}
  {\path{doi:10.1103/PhysRevB.36.840}}.

\bibitem{Svendlindh1989}
P.~Svedlindh, K.~Niskanen, P.~Norling, P.~Nordblad, L.~Lundgren, B.~Lönnberg,
  T.~Lundström, {Anti-Meissner effect in the BiSrCaCuO-system}, Physica C:
  Superconductivity and its Applications 162-164 (1989) 1365 -- 1366.
\newblock \href {https://doi.org/https://doi.org/10.1016/0921-4534(89)90735-1}
  {\path{doi:https://doi.org/10.1016/0921-4534(89)90735-1}}.

\bibitem{Geim1998}
A.~K. Geim, S.~V. Dubonos, J.~G.~S. Lok, M.~Henini, J.~C. Maan, {Paramagnetic
  Meissner effect in small superconductors}, Nature 396~(6707) (1998) 144--146.
\newblock \href {https://doi.org/10.1038/24110} {\path{doi:10.1038/24110}}.

\bibitem{Palacios1998}
J.~J. Palacios, {Vortex matter in superconducting mesoscopic disks: Structure,
  magnetization, and phase transitions}, Phys. Rev. B 58 (1998) R5948--R5951.
\newblock \href {https://doi.org/10.1103/PhysRevB.58.R5948}
  {\path{doi:10.1103/PhysRevB.58.R5948}}.

\bibitem{AraujoMoreira1997}
F.~M. Araujo-Moreira, P.~Barbara, A.~B. Cawthorne, C.~J. Lobb, {Reentrant ac
  Magnetic Susceptibility in Josephson-Junction Arrays}, Phys. Rev. Lett. 78
  (1997) 4625--4628.
\newblock \href {https://doi.org/10.1103/PhysRevLett.78.4625}
  {\path{doi:10.1103/PhysRevLett.78.4625}}.

\bibitem{Barbara1999}
P.~Barbara, F.~M. Araujo-Moreira, A.~B. Cawthorne, C.~J. Lobb, {Reentrant ac
  magnetic susceptibility in Josephson-junction arrays: An alternative
  explanation for the paramagnetic Meissner effect}, Phys. Rev. B 60 (1999)
  7489--7495.
\newblock \href {https://doi.org/10.1103/PhysRevB.60.7489}
  {\path{doi:10.1103/PhysRevB.60.7489}}.

\bibitem{Chu2006}
S.~Chu, A.~J. Schwartz, T.~B. Massalski, D.~E. Laughlin, {Extrinsic
  paramagnetic Meissner effect in multiphase indium-tin alloys}, Applied
  Physics Letters 89~(11) (2006) 111903.
\newblock \href {https://doi.org/10.1063/1.2352805}
  {\path{doi:10.1063/1.2352805}}.

\bibitem{SigristRice1995}
M.~Sigrist, T.~M. Rice, {Unusual paramagnetic phenomena in granular
  high-temperature superconductors---A consequence of $d$- wave pairing?}, Rev.
  Mod. Phys. 67 (1995) 503--513.
\newblock \href {https://doi.org/10.1103/RevModPhys.67.503}
  {\path{doi:10.1103/RevModPhys.67.503}}.

\bibitem{KoshelevLarkin}
A.~E. Koshelev, A.~I. Larkin, {Paramagnetic moment in field-cooled
  superconducting plates: Paramagnetic Meissner effect}, Phys. Rev. B 52 (1995)
  13559--13562.
\newblock \href {https://doi.org/10.1103/PhysRevB.52.13559}
  {\path{doi:10.1103/PhysRevB.52.13559}}.

\bibitem{Zharkov2001}
G.~F. Zharkov, {Paramagnetic Meissner effect in superconductors from
  self-consistent solution of Ginzburg-Landau equations}, Phys. Rev. B 63
  (2001) 214502.
\newblock \href {https://doi.org/10.1103/PhysRevB.63.214502}
  {\path{doi:10.1103/PhysRevB.63.214502}}.

\bibitem{Johnston2013}
D.~C. Johnston, {Elaboration of the $\alpha$-model derived from the {BCS}
  theory of superconductivity}, Superconductor Science and Technology 26~(11)
  (2013) 115011.
\newblock \href {https://doi.org/10.1088/0953-2048/26/11/115011}
  {\path{doi:10.1088/0953-2048/26/11/115011}}.

\bibitem{Huang2006}
C.~L. Huang, J.-Y. Lin, C.~P. Sun, T.~K. Lee, J.~D. Kim, E.~M. Choi, S.~I. Lee,
  H.~D. Yang, {Comparative analysis of specific heat of
  ${\mathrm{YNi}}_{2}{\mathrm{B}}_{2}\mathrm{C}$ using nodal and two-gap
  models}, Phys. Rev. B 73 (2006) 012502.
\newblock \href {https://doi.org/10.1103/PhysRevB.73.012502}
  {\path{doi:10.1103/PhysRevB.73.012502}}.

\bibitem{Custers2003}
J.~Custers, P.~Gegenwart, H.~Wilhelm, K.~Neumaier, Y.~Tokiwa, O.~Trovarelli,
  C.~Geibel, F.~Steglich, C.~P{\'e}pin, P.~Coleman, {The break-up of heavy
  electrons at a quantum critical point}, Nature 424~(6948) (2003) 524--527.
\newblock \href {https://doi.org/10.1038/nature01774}
  {\path{doi:10.1038/nature01774}}.

\bibitem{Bianchi2003}
A.~Bianchi, R.~Movshovich, I.~Vekhter, P.~G. Pagliuso, J.~L. Sarrao, {Avoided
  Antiferromagnetic Order and Quantum Critical Point in
  $\mathrm{C}\mathrm{e}\mathrm{C}\mathrm{o}\mathrm{I}{\mathrm{n}}_{\mathrm{5}}$},
  Phys. Rev. Lett. 91 (2003) 257001.
\newblock \href {https://doi.org/10.1103/PhysRevLett.91.257001}
  {\path{doi:10.1103/PhysRevLett.91.257001}}.

\bibitem{Sidorov2002}
V.~A. Sidorov, M.~Nicklas, P.~G. Pagliuso, J.~L. Sarrao, Y.~Bang, A.~V.
  Balatsky, J.~D. Thompson, {Superconductivity and Quantum Criticality in
  $\mathrm{C}\mathrm{e}\mathrm{C}\mathrm{o}\mathrm{I}{\mathrm{n}}_{\mathrm{5}}$},
  Phys. Rev. Lett. 89 (2002) 157004.
\newblock \href {https://doi.org/10.1103/PhysRevLett.89.157004}
  {\path{doi:10.1103/PhysRevLett.89.157004}}.

\bibitem{Analytis2014}
J.~G. Analytis, H.-H. Kuo, R.~D. McDonald, M.~Wartenbe, P.~M.~C. Rourke, N.~E.
  Hussey, I.~R. Fisher, {Transport near a quantum critical point in
  BaFe$_2$(As$_{1-x}$P$_x$)$_2$}, Nature Physics 10~(3) (2014) 194--197.
\newblock \href {https://doi.org/10.1038/nphys2869}
  {\path{doi:10.1038/nphys2869}}.

\bibitem{Grinenko2017}
V.~Grinenko, K.~Iida, F.~Kurth, D.~V. Efremov, S.-L. Drechsler,
  I.~Cherniavskii, I.~Morozov, J.~H{\"a}nisch, T.~F{\"o}rster, C.~Tarantini,
  J.~Jaroszynski, B.~Maiorov, M.~Jaime, A.~Yamamoto, I.~Nakamura, R.~Fujimoto,
  T.~Hatano, H.~Ikuta, R.~H{\"u}hne, {Selective mass enhancement close to the
  quantum critical point in BaFe$_2$(As$_{1-x}$P$_x$)$_2$}, Scientific Reports
  7~(1) (2017) 4589.
\newblock \href {https://doi.org/10.1038/s41598-017-04724-3}
  {\path{doi:10.1038/s41598-017-04724-3}}.

\bibitem{Nakajima2007}
Y.~Nakajima, H.~Shishido, H.~Nakai, T.~Shibauchi, K.~Behnia, K.~Izawa, M.~Hedo,
  Y.~Uwatoko, T.~Matsumoto, R.~Settai, Y.~Ōnuki, H.~Kontani, Y.~Matsuda,
  {Non-Fermi Liquid Behavior in the Magnetotransport of CeMIn5 (M: Co and Rh):
  Striking Similarity between Quasi Two-Dimensional Heavy Fermion and High-Tc
  Cuprates}, Journal of the Physical Society of Japan 76~(2) (2007) 024703.
\newblock \href {https://doi.org/10.1143/JPSJ.76.024703}
  {\path{doi:10.1143/JPSJ.76.024703}}.

\bibitem{Valla1999}
T.~Valla, A.~V. Fedorov, P.~D. Johnson, B.~O. Wells, S.~L. Hulbert, Q.~Li,
  G.~D. Gu, N.~Koshizuka, {Evidence for Quantum Critical Behavior in the
  Optimally Doped Cuprate Bi$_2$Sr$_2$CaCu$_2$O$_{8+\delta}$}, Science
  285~(5436) (1999) 2110--2113.
\newblock \href {https://doi.org/10.1126/science.285.5436.2110}
  {\path{doi:10.1126/science.285.5436.2110}}.

\bibitem{Keimer2015}
B.~Keimer, S.~A. Kivelson, M.~R. Norman, S.~Uchida, J.~Zaanen, {From quantum
  matter to high-temperature superconductivity in copper oxides}, Nature
  518~(7538) (2015) 179--186.
\newblock \href {https://doi.org/10.1038/nature14165}
  {\path{doi:10.1038/nature14165}}.

\bibitem{Sachdev2010}
S.~Sachdev, {Where is the quantum critical point in the cuprate
  superconductors?}, Physica Status Solidi (B) 247~(3) (2010) 537--543.
\newblock \href {https://doi.org/10.1002/pssb.200983037}
  {\path{doi:10.1002/pssb.200983037}}.

\bibitem{Zhou2004}
Z.~X. Zhou, S.~McCall, C.~S. Alexander, J.~E. Crow, P.~Schlottmann, A.~Bianchi,
  C.~Capan, R.~Movshovich, K.~H. Kim, M.~Jaime, N.~Harrison, M.~K. Haas, R.~J.
  Cava, G.~Cao, {Transport and thermodynamic properties of
  ${\mathrm{Sr}}_{3}{\mathrm{Ru}}_{2}{\mathrm{O}}_{7}$ near the quantum
  critical point}, Phys. Rev. B 69 (2004) 140409.
\newblock \href {https://doi.org/10.1103/PhysRevB.69.140409}
  {\path{doi:10.1103/PhysRevB.69.140409}}.

\bibitem{Grigera2004}
S.~A. Grigera, P.~Gegenwart, R.~A. Borzi, F.~Weickert, A.~J. Schofield, R.~S.
  Perry, T.~Tayama, T.~Sakakibara, Y.~Maeno, A.~G. Green, A.~P. Mackenzie,
  {Disorder-Sensitive Phase Formation Linked to Metamagnetic Quantum
  Criticality}, Science 306~(5699) (2004) 1154--1157.
\newblock \href {https://doi.org/10.1126/science.1104306}
  {\path{doi:10.1126/science.1104306}}.

\bibitem{Liu2020}
C.~Liu, V.~F.~C. Humbert, T.~M. Bretz-Sullivan, G.~Wang, D.~Hong, F.~Wrobel,
  J.~Zhang, J.~D. Hoffman, J.~E. Pearson, J.~S. Jiang, C.~Chang, A.~Suslov,
  N.~Mason, M.~R. Norman, A.~Bhattacharya, {Observation of an antiferromagnetic
  quantum critical point in high-purity LaNiO$_3$}, Nature Communications
  11~(1) (2020) 1402.
\newblock \href {https://doi.org/10.1038/s41467-020-15143-w}
  {\path{doi:10.1038/s41467-020-15143-w}}.

\bibitem{Das2019}
D.~Das, D.~Gnida, P.~Wi{\'s}niewski, D.~Kaczorowski, {Magnetic field-driven
  quantum criticality in antiferromagnetic CePtIn$_4$}, Proceedings of the
  National Academy of Sciences 116~(41) (2019) 20333--20338.
\newblock \href {https://doi.org/10.1073/pnas.1910293116}
  {\path{doi:10.1073/pnas.1910293116}}.

\bibitem{Sachdev_2011}
S.~Sachdev, {Quantum Phase Transitions}, 2nd Edition, Cambridge University
  Press, 2011.
\newblock \href {https://doi.org/10.1017/CBO9780511973765}
  {\path{doi:10.1017/CBO9780511973765}}.

\bibitem{Kordyuk2015}
A.~A. Kordyuk, {Pseudogap from ARPES experiment: Three gaps in cuprates and
  topological superconductivity (Review Article)}, Low Temperature Physics
  41~(5) (2015) 319--341.
\newblock \href {https://doi.org/10.1063/1.4919371}
  {\path{doi:10.1063/1.4919371}}.

\bibitem{Cyr-Choiniere2018}
O.~Cyr-Choini\`ere, R.~Daou, F.~Lalibert\'e, C.~Collignon, S.~Badoux,
  D.~LeBoeuf, J.~Chang, B.~J. Ramshaw, D.~A. Bonn, W.~N. Hardy, R.~Liang, J.-Q.
  Yan, J.-G. Cheng, J.-S. Zhou, J.~B. Goodenough, S.~Pyon, T.~Takayama,
  H.~Takagi, N.~Doiron-Leyraud, L.~Taillefer, {Pseudogap temperature ${T}^{*}$
  of cuprate superconductors from the Nernst effect}, Phys. Rev. B 97 (2018)
  064502.
\newblock \href {https://doi.org/10.1103/PhysRevB.97.064502}
  {\path{doi:10.1103/PhysRevB.97.064502}}.

\bibitem{Calegari2015}
E.~J. Calegari, A.~C. Lausmann, S.~G. Magalhaes, C.~M. Chaves, A.~Troper,
  {Pseudogap and the specific heat of high Tcsuperconductors: a Hubbard model
  in a n-pole approximation}, Journal of Physics: Conference Series 592 (2015)
  012075.
\newblock \href {https://doi.org/10.1088/1742-6596/592/1/012075}
  {\path{doi:10.1088/1742-6596/592/1/012075}}.

\bibitem{Agterberg2020}
D.~F. Agterberg, J.~S. Davis, S.~D. Edkins, E.~Fradkin, D.~J. Van~Harlingen,
  S.~A. Kivelson, P.~A. Lee, L.~Radzihovsky, J.~M. Tranquada, Y.~Wang, {The
  Physics of Pair-Density Waves: Cuprate Superconductors and Beyond}, Annual
  Review of Condensed Matter Physics 11~(1) (2020) 231--270.
\newblock \href {https://doi.org/10.1146/annurev-conmatphys-031119-050711}
  {\path{doi:10.1146/annurev-conmatphys-031119-050711}}.

\bibitem{Choubey2020}
P.~Choubey, S.~H. Joo, K.~Fujita, Z.~Du, S.~D. Edkins, M.~H. Hamidian,
  H.~Eisaki, S.~Uchida, A.~P. Mackenzie, J.~Lee, J.~C.~S. Davis, P.~J.
  Hirschfeld, Atomic-scale electronic structure of the cuprate pair density
  wave state coexisting with superconductivity, Proceedings of the National
  Academy of Sciences (2020).
\newblock \href {https://doi.org/10.1073/pnas.2002429117}
  {\path{doi:10.1073/pnas.2002429117}}.

\bibitem{Badoux2016}
S.~Badoux, W.~Tabis, F.~Lalibert{\'e}, G.~Grissonnanche, B.~Vignolle,
  D.~Vignolles, J.~B{\'e}ard, D.~A. Bonn, W.~N. Hardy, R.~Liang,
  N.~Doiron-Leyraud, L.~Taillefer, C.~Proust, {Change of carrier density at the
  pseudogap critical point of a cuprate superconductor}, Nature 531~(7593)
  (2016) 210--214.
\newblock \href {https://doi.org/10.1038/nature16983}
  {\path{doi:10.1038/nature16983}}.

\bibitem{LeBoeuf2011}
D.~LeBoeuf, N.~Doiron-Leyraud, B.~Vignolle, M.~Sutherland, B.~J. Ramshaw,
  J.~Levallois, R.~Daou, F.~Lalibert\'e, O.~Cyr-Choini\`ere, J.~Chang, Y.~J.
  Jo, L.~Balicas, R.~Liang, D.~A. Bonn, W.~N. Hardy, C.~Proust, L.~Taillefer,
  {Lifshitz critical point in the cuprate superconductor
  ${\mathrm{YBa}}_{2}{\mathrm{Cu}}_{3}{\mathrm{O}}_{y}$ from high-field Hall
  effect measurements}, Phys. Rev. B 83 (2011) 054506.
\newblock \href {https://doi.org/10.1103/PhysRevB.83.054506}
  {\path{doi:10.1103/PhysRevB.83.054506}}.

\bibitem{Doiron-Leyraud2015}
N.~Doiron-Leyraud, S.~Badoux, S.~Ren{\'e}~de Cotret, S.~Lepault, D.~LeBoeuf,
  F.~Lalibert{\'e}, E.~Hassinger, B.~J. Ramshaw, D.~A. Bonn, W.~N. Hardy,
  R.~Liang, J.-H. Park, D.~Vignolles, B.~Vignolle, L.~Taillefer, C.~Proust,
  {Evidence for a small hole pocket in the Fermi surface of underdoped
  YBa$_2$Cu$_3$O$_y$}, Nature Communications 6~(1) (2015) 6034.
\newblock \href {https://doi.org/10.1038/ncomms7034}
  {\path{doi:10.1038/ncomms7034}}.

\bibitem{Laliberte2011}
F.~Lalibert{\'e}, J.~Chang, N.~Doiron-Leyraud, E.~Hassinger, R.~Daou,
  M.~Rondeau, B.~J. Ramshaw, R.~Liang, D.~A. Bonn, W.~N. Hardy, S.~Pyon,
  T.~Takayama, H.~Takagi, I.~Sheikin, L.~Malone, C.~Proust, K.~Behnia,
  L.~Taillefer, {Fermi-surface reconstruction by stripe order in cuprate
  superconductors}, Nature Communications 2~(1) (2011) 432.
\newblock \href {https://doi.org/10.1038/ncomms1440}
  {\path{doi:10.1038/ncomms1440}}.

\bibitem{Haug2010}
D.~Haug, V.~Hinkov, Y.~Sidis, P.~Bourges, N.~B. Christensen, A.~Ivanov,
  T.~Keller, C.~T. Lin, B.~Keimer, {Neutron scattering study of the magnetic
  phase diagram of underdoped YBa$_2$Cu$_3$O$_{6+x}$}, New Journal of Physics
  12~(10) (2010) 105006.
\newblock \href {https://doi.org/10.1088/1367-2630/12/10/105006}
  {\path{doi:10.1088/1367-2630/12/10/105006}}.

\bibitem{Presland1991}
M.~Presland, J.~Tallon, R.~Buckley, R.~Liu, N.~Flower, {General trends in
  oxygen stoichiometry effects on Tc in Bi and Tl superconductors}, Physica C:
  Superconductivity 176~(1) (1991) 95 -- 105.

\bibitem{Hucker2014}
M.~H\"ucker, N.~B. Christensen, A.~T. Holmes, E.~Blackburn, E.~M. Forgan,
  R.~Liang, D.~A. Bonn, W.~N. Hardy, O.~Gutowski, M.~v. Zimmermann, S.~M.
  Hayden, J.~Chang, {Competing charge, spin, and superconducting orders in
  underdoped ${\mathrm{YBa}}_{2}{\mathrm{Cu}}_{3}{\mathrm{O}}_{y}$}, Phys. Rev.
  B 90 (2014) 054514.
\newblock \href {https://doi.org/10.1103/PhysRevB.90.054514}
  {\path{doi:10.1103/PhysRevB.90.054514}}.

\bibitem{Blanco2014}
S.~Blanco-Canosa, A.~Frano, E.~Schierle, J.~Porras, T.~Loew, M.~Minola,
  M.~Bluschke, E.~Weschke, B.~Keimer, M.~Le~Tacon, {Resonant x-ray scattering
  study of charge-density wave correlations in
  ${\mathrm{YBa}}_{2}{\mathrm{Cu}}_{3}{\mathrm{O}}_{6+x}$}, Phys. Rev. B 90
  (2014) 054513.
\newblock \href {https://doi.org/10.1103/PhysRevB.90.054513}
  {\path{doi:10.1103/PhysRevB.90.054513}}.

\bibitem{Coleman2005}
P.~Coleman, A.~J. Schofield, Quantum criticality, Nature 433~(7023) (2005)
  226--229.
\newblock \href {https://doi.org/10.1038/nature03279}
  {\path{doi:10.1038/nature03279}}.

\bibitem{Degen2014}
T.~Degen, M.~Sadki, E.~Bron, U.~K\"onig, G.~N\'enert, {The HighScore suite},
  Powder Diffraction 29~(S2) (2014) S13 -- S18.
\newblock \href {https://doi.org/10.1017/S0885715614000840}
  {\path{doi:10.1017/S0885715614000840}}.

\bibitem{Petricek2014}
V.~Pet\u{r}\'i\u{c}ek, M.~Du\u{s}ek, L.~Palatinus, {Crystallographic Computing
  System JANA2006: General features}, Zeitschrift f\"ur Kristallographie -
  Crystalline Materials 229~(5) (2014) 345--352.
\newblock \href {https://doi.org/doi:10.1515/zkri-2014-1737}
  {\path{doi:doi:10.1515/zkri-2014-1737}}.

\bibitem{Rietveld1967}
H.~M. Rietveld, {Line profiles of neutron powder-diffraction peaks for
  structure refinement}, Acta Crystallographica 22~(1) (1967) 151--152.
\newblock \href {https://doi.org/10.1107/S0365110X67000234}
  {\path{doi:10.1107/S0365110X67000234}}.

\bibitem{Rietveld1969}
H.~M. Rietveld, {A profile refinement method for nuclear and magnetic
  structures}, Journal of Applied Crystallography 2~(2) (1969) 65--71.
\newblock \href {https://doi.org/10.1107/S0021889869006558}
  {\path{doi:10.1107/S0021889869006558}}.

\bibitem{Lebail1988}
A.~{Le Bail}, H.~Duroy, J.~Fourquet, {Ab-initio structure determination of
  LiSbWO$_6$ by X-ray powder diffraction}, Materials Research Bulletin 23~(3)
  (1988) 447--452.
\newblock \href {https://doi.org/https://doi.org/10.1016/0025-5408(88)90019-0}
  {\path{doi:https://doi.org/10.1016/0025-5408(88)90019-0}}.

\bibitem{Xie2012}
L.~Xie, T.~Su, X.~Li, {Magnetic field dependence of Schottky anomaly in the
  specific heats of stripe-ordered superconductors
  La$_{1.6-x}$Nd$_{0.4}$Sr$_x$CuO$_4$}, Physica C: Superconductivity 480 (2012)
  14 -- 18.

\bibitem{BaSeO4}
We have estimated the $C/T$ of this material using temperature dependence of
  BaSO$_{4}$ and the fact that at room temperature the standard molar entropy
  132.2 J/(mol K) of it corresponds to $C_{BaSO_{4}}=$101.8 J/(mol K) (CRC
  Handbook of Chemistry and Physics, 95 edition, CRC 2014, pp.4-50). At the
  same temperature, the standard molar entropy of BaSeO$_{4}$ is 134 J/(mol K)
  (http://chemister.ru/Database/properties-en.php?dbid=1\&id=7158), so that its
  heat capacity should not be much different from BaSO$_{4}$.

\bibitem{Matskevich-arxiv}
N.~I. Matskevich, Y.~F. Minenkov, G.~A. Berezovskii,
  \href{https://arxiv.org/abs/1401.7422}{Thermodynamic properties of yttrium
  cuprate}, arXiv.1401.7422 (2014).
\newline\urlprefix\url{https://arxiv.org/abs/1401.7422}

\bibitem{Eckert1988}
D.~Eckert, A.~Junod, T.~Graf, J.~Muller, {Low Temperature specific heat of
  YBa$_3$Cu$_2$O$_7$, Y$_2$BaCuO$_5$, CuO and BaCuO$_{2+x}$}, Physica C:
  Superconductivity 153-155 (1988) 1038 -- 1039.

\bibitem{Lin2011}
J.-Y. Lin, Y.~S. Hsieh, D.~A. Chareev, A.~N. Vasiliev, Y.~Parsons, H.~D. Yang,
  {Coexistence of isotropic and extended $s$-wave order parameters in FeSe as
  revealed by low-temperature specific heat}, Phys. Rev. B 84 (2011) 220507.
\newblock \href {https://doi.org/10.1103/PhysRevB.84.220507}
  {\path{doi:10.1103/PhysRevB.84.220507}}.

\bibitem{Carless1983}
D.~C. Carless, H.~E. Hall, J.~R. Hook, {Vibrating wire measurements in
  liquid3He II. The superfluid B phase}, Journal of Low Temperature Physics
  50~(5) (1983) 605--633.
\newblock \href {https://doi.org/10.1007/BF00683498}
  {\path{doi:10.1007/BF00683498}}.

\bibitem{Collocott88}
S.~J. Collocott, R.~Driver, L.~Dale, S.~X. Dou, {Schottky anomaly in the heat
  capacity of the high-${T}_{c}$ superconductor
  ${\mathrm{YBa}}_{2}$${\mathrm{Cu}}_{3}$${\mathrm{O}}_{7}$}, Phys. Rev. B 37
  (1988) 7917--7919.
\newblock \href {https://doi.org/10.1103/PhysRevB.37.7917}
  {\path{doi:10.1103/PhysRevB.37.7917}}.

\bibitem{Mu2007}
G.~Mu, Y.~Wang, L.~Shan, H.-H. Wen, {Possible nodeless superconductivity in the
  noncentrosymmetric superconductor
  ${\mathrm{Mg}}_{12\ensuremath{-}\ensuremath{\delta}}{\mathrm{Ir}}_{19}{\mathrm{B}}_{16}$},
  Phys. Rev. B 76 (2007) 064527.
\newblock \href {https://doi.org/10.1103/PhysRevB.76.064527}
  {\path{doi:10.1103/PhysRevB.76.064527}}.

\end{thebibliography}





\end{document}